\documentclass[10pt]{article}
\usepackage[square,authoryear]{natbib}
\usepackage{graphicx}
\usepackage{marsden_article}

\usepackage{epstopdf,framed}
\usepackage{pdfsync}
\begin{document}
\newtheorem{remark}[theorem]{Remark}

\title{A variational derivation of the nonequilibrium thermodynamics of a moist atmosphere \textcolor{black}{with rain process and its pseudoincompressible approximation}}
\vspace{-0.2in}

%%% TodoFGB
\newcommand{\todoFGB}[1]{\vspace{5 mm}\par \noindent
\framebox{\begin{minipage}[c]{0.95 \textwidth} \color{red}FGB: \tt #1
\end{minipage}}\vspace{5 mm}\par}
%%%

%\author{ Fran\c{c}ois Gay-Balmaz and
%Hiroaki Yoshimura 
%\date{Sep 27, 2010}
%\date{Jan 9, 2011}
%\date{July 10, 2011}
%}

\author{Fran\c{c}ois Gay-Balmaz$^{1}$}
%\author{
%\hspace{-1cm}
%\begin{tabular}{c}
%Fran\c{c}ois Gay-Balmaz$^{1}$\\ 
%LMD/IPSL, CNRS, Ecole normale sup\'erieure, PSL Research University\\
%Ecole polytechnique, Universit\'e Paris-Saclay, Sorbonne Universit\'es, UPMC Univ Paris 06\\
%24 Rue Lhomond 75005 Paris, France\\
%francois.gay-balmaz@lmd.ens.fr\\
%\end{tabular}\\
%}

\date{}

\addtocounter{footnote}{1}
\footnotetext{LMD/IPSL, CNRS, Ecole normale sup\'erieure, PSL Research University,
Ecole polytechnique, Universit\'e Paris-Saclay, Sorbonne Universit\'es, UPMC Univ Paris 06,
24 Rue Lhomond 75005 Paris, France,
francois.gay-balmaz@lmd.ens.fr}

\maketitle

\begin{center}
\abstract{Irreversible processes play a major role in the description and prediction of atmospheric dynamics.
 In this paper, we present a variational derivation of the evolution equations for a moist atmosphere with rain process and subject to the irreversible processes of viscosity, heat conduction, diffusion, and phase transition. This derivation is based on a general variational formalism for nonequilibrium thermodynamics which extends Hamilton's principle to incorporates irreversible processes. It is valid for any state equation and thus also covers the case of the atmosphere of other planets. In this approach, the second law of thermodynamics is understood as a nonlinear constraint formulated with the help of new variables, called thermodynamic displacements, whose time derivative coincides with the thermodynamic force of the irreversible process. The formulation is written both in the Lagrangian and Eulerian descriptions and can be directly adapted to oceanic dynamics. \textcolor{black}{We illustrate the efficiency of our variational formulation as a modeling tool in atmospheric thermodynamics, by deriving a pseudoincompressible model for moist atmospheric thermodynamics with general equations of state and subject to the irreversible processes of viscosity, heat conduction, diffusion, and phase transition.}}
\vspace{2mm}

\end{center}
%\tableofcontents

\section{Introduction}

The partial differential equations governing the thermodynamics of the atmosphere are of obvious importance for weather and climate prediction. These equations are well-known for their extreme complexity, both from the theoretical and the computational side, which is in part due to the many physical processes that they involve, such as phase transition, cloud formation, precipitation, and radiation.

\medskip

In absence of irreversible processes, the equations of atmospheric dynamic can be derived by applying Hamilton's variational principle to the Lagrangian function of the fluid. This is in agreement with a fundamental fact from classical reversible mechanics, namely that the motion of the mechanical system is governed by the Euler-Lagrange equations which, in turn, describe the critical points of the action functional of this Lagrangian among all possible trajectories with prescribed values at the temporal extremities.
Hamilton's principle for fluid mechanics in the Lagrangian description has been discussed at least since the works of \cite{He1955}, for an incompressible fluid and \cite{Se1959} and \cite{Ec1960} for compressible flows, see also \cite{TrTo1960} for further references. While in the Lagrangian description this principle is a straightforward extension of the Hamilton principle of particles mechanics, in the Eulerian description the variational principle is much more involved and several approaches have been developed, see \cite{Li1963}, \cite{SeWh1968}, \cite{Br1970}. We refer to \cite{Sa1983} and \cite{Sa1988} for further developments in the context of geophysical fluids, see also \cite{Mu1995}. In \cite{HoMaRa2002}, the variational principle in Eulerian description is obtained via the Euler-Poincar\'e reduction theory for several geophysical fluid models, by exploiting the relabelling symmetries. Physically, the existence of a variational Hamilton's principle manifests itself in specific Lagrangian conservation laws, the most celebrated being the conservation of potential vorticity which
alone allows one to understand many processes taking place in the atmosphere, the
ocean, or in laboratory experiments (e.g. \cite{HoMIRo1985}). Conservation of potential vorticity is related to Kelvin's circulation theorem which, thanks to variational principle, can be interpreted in terms of the general Noether theorem, linking conservation laws to symmetries. More recently, the variational formulation of geophysical fluids has been crucially used in \cite{DeGBGaZe2014} and \cite{BaGB2017a} to derive structure preserving numerical schemes for rotating stratified fluids in the Boussinesq and pseudo-incompressible approximations, following \cite{PaMuToKaMaDe2010}, \cite{GaMuPaMaDe2011}. Extension to the compressible case was developed in \cite{BaGB2017b}.

\medskip

Irreversible processes such as phase transition, cloud formation, precipitation, and radiation, play a major role in the description and prediction of atmospheric dynamics and it has been a long standing 
question whether or not the variational formulations mentioned above can be extended to include all these irreversible processes.
In this paper, we shall positively answer this question by presenting a variational derivation for the thermodynamic of a moist atmosphere that includes the irreversible processes of viscosity, heat conduction, diffusion, and phase transition. We shall use the variational formulation of nonequilibrium thermodynamics recently developed in \cite{GBYo2016a,GBYo2016b}. The main aspect of this approach is the introduction, for each of the irreversible process, of a new variable, called the \textit{thermodynamic displacement}, whose time derivative coincides with the thermodynamic force of the process. Recall that, according to \cite{deGrootMazur1969}, the thermodynamic force of a process is related to the non-uniformity of the system (the gradient of the temperature for instance) or to the deviation of some internal state variables from their equilibrium values (the chemical affinity for instance). It turns out that the entropy source of the system is a sum of terms, each being the product of a thermodynamic flux (or flow)$^{1}$ characterizing an irreversible process, and a thermodynamic force, see \cite{deGrootMazur1969}.\addtocounter{footnote}{1}
\footnotetext{In \cite{deGrootMazur1969} both terminologies \textit{thermodynamic flux} and \textit{thermodynamic flow} are used for the same concept.}
As we will explain below, the thermodynamic displacements introduced in \cite{GBYo2016a,GBYo2016b} allow the formulation of the second law of thermodynamics as a nonlinear constraint to be used in the variational formulation, both for the Lagrangian and Eulerian descriptions. This constraint involves the phenomenological expression of each of the thermodynamic fluxes (or flows) in terms of the thermodynamic forces. It is hence called the phenomenological constraint. From the point of view of atmospheric modelling, this constraint encodes the various parametrizations of subgrid scale effects in general circulation models of the atmosphere.
In absence of irreversible processes, the constraint disappears and our variational formulation consistently recovers the classical variational principles in Lagrangian and Eulerian descriptions.

%Determining the equations and solving them is crucial for weather and climate prediction ....

\medskip

The atmosphere of the Earth is composed of dry air, water substance in any of its three phases, and atmospheric aerosols. 
For practical use in meteorology, dry air, whose main components are Nitrogen and Oxygen, can be regarded as an ideal gas. 
Unlike the components of dry air, the proportion of the gas phase of water, i.e., water vapor, is very variable and plays a major role in the thermodynamics of the atmosphere because of its ability to condense under atmospheric conditions.
The condensed phases of water consist of cloud particles (water droplet or ice crystal) that follow the motion of dry air and of hydrometeors (such as rain and snow) that are falling through the air. The atmospheric aerosols are solid and liquid particles in suspension (other than that of water substance) whose study is very important for atmospheric chemistry, cloud and precipitation physics, and for atmospheric radiation and optics. It is not significant for atmospheric thermodynamics, and shall not be considered here.
In the first part of the paper (Section \ref{1_dry}), we restrict our approach to the case of dry air. In this case, the only irreversible processes are viscosity and heat conduction. This situation allows us to introduce the main ideas of our approach in a simplified context. In the second part of the paper (Section \ref{2_moist}) we consider the case of a moist atmosphere by including water substance in its different phases, namely, water vapor, liquid water, and solid water. The irreversible processes considered are viscosity, heat conduction, vapor diffusion, and phase changes, as well as rain process. We also compute the impact of the irreversible processes on the potential vorticity equation and on Kelvin's circulation theorem, by staying in the general Lagrangian framework, which provides a useful unified treatment of potential vorticity and circulation theorems for various approximations of the equations of atmospheric dynamics. \textcolor{black}{In order to illustrate the efficiency of our variational formulation as a modeling tool in atmospheric thermodynamics, we derive in Section \ref{Sec_PI} a thermodynamically consistent pseudoincompressible model for moist atmospheric thermodynamics with general equations of state and subject to the irreversible processes of viscosity, heat conduction, diffusion, and phase transition.}
Finally, in Section \ref{3_ocean}, we quickly mention that our approach directly applies to oceanic dynamics with the irreversible process of viscosity, heat conduction, and salt diffusion included.

\section{Variational derivation of the thermodynamic of a dry atmosphere}\label{1_dry} 

In this section we consider the dynamics of a dry atmosphere subject to the irreversible processes of viscosity and heat conduction.

The equation of state of dry air is the ideal gas law
\begin{equation}\label{state_equation}
v=v(p,T)= \frac{R _d T}{p}, 
\end{equation} 
where $v$ is the specific volume, $p$ is the pressure, $T$ is the temperature, $R _d=R ^\ast/m_d$  is the gas constant for dry air written in terms of the universal gas constant $R ^\ast $ and the mean molecular weight of dry air $m_d$. The expression of all other thermodynamic variables for dry air in terms of $p$ and $T$ can be derived by using the 
equation of state \eqref{state_equation} and the fact that the specific heat $C_p$ at constant pressure can be assumed to be constant in the atmosphere, which is the hypothesis for a perfect gas. For example, it is deduced that the specific internal energy and the specific entropy are
\[
u= C_vT, \quad \eta = C_p\ln T-R_d\ln p+  \eta  _0,
\]
where $C_v$ is the specific heat at constant volume, $ \eta _0$ is a constant, and we assume that the internal energy at $T=0$ K is zero.

Our variational formulation is more naturally expressed in terms of \textit{density} variables rather than \textit{specific} variables, hence we will write the Lagrangian in terms of the mass density $ \rho= \frac{1}{v}$ and entropy density $s = \frac{ \eta }{v}$. In addition, the variational derivation has a simpler form in Lagrangian description, this is why we shall first write it in Lagrangian variables and then later deduce from it the Eulerian form.

Let us denote by $ \mathbf{x} = \varphi (t, \mathbf{X} )$ the Lagrangian trajectory of dry air particles. The variable $ \mathbf{X} $ refers to the label of the particle and $ \mathbf{x} $ is its current location. The Lagrangian and Eulerian wind velocities $ \dot\varphi = \frac{d}{dt} \varphi $ and $ \mathbf{v} $ are related as
\[
\frac{d}{dt} \varphi (t, \mathbf{X} )= \mathbf{v} (t,\varphi (t, \mathbf{X} )).
\]
We shall denote by $S(t, \mathbf{X} )$ the entropy density in Lagrangian representation, related to $s( t, \mathbf{x} )$ as $S(t, \mathbf{X} )= s(t, \varphi (t, \mathbf{X} ))| \nabla  \varphi (t, \mathbf{X} )| $, where $| \nabla  \varphi |$ denotes the Jacobian of $ \varphi $.
The mass density in Lagrangian representation is $ \varrho  _0(\mathbf{X} )=  \rho (t, \varphi (t, \mathbf{X} ))| \nabla  \varphi (t, \mathbf{X} )|$.
Note that because of mass conservation, the mass density $ \varrho _0$ in Lagrangian representation is time independent. This is in contrast with the time dependence of $S$, due to the presence of the irreversible processes.

The Lagrangian of the dry atmosphere consists of the sum of the kinetic energy and of the contribution of Earth rotation, to which is substracted the gravitational potential $\Phi  $ and the internal energy $u$. In terms of the Lagrangian variables $ \varphi $, $\dot\varphi $, and $S$, it reads
\begin{equation}\label{Lagrangian} 
L( \varphi , \dot \varphi , S)= \int_ \mathcal{D}  \varrho  _0 \left[ \frac{1}{2} \left| \dot\varphi\right| ^2 + \mathbf{R}  (\varphi ) \cdot \dot \varphi  -\Phi ( \varphi ) - u  \left(  \frac{S}{ \varrho _0} ,\frac{|  \nabla \varphi |}{\varrho _0} \right)\right] d \mathbf{X} =: \int_ \mathcal{D} \mathfrak{L}\,d \mathbf{X} .
\end{equation} 
Here the vector field $ \mathbf{R} $ is the vector potential for the Coriolis parameter, i.e., $ \operatorname{curl} \mathbf{R} =2 \boldsymbol{\Omega} $, where $ \boldsymbol{\Omega} $ is the angular velocity of the Earth. On the right hand side of \eqref{Lagrangian} we defined the Lagrangian density $ \mathfrak{L} $ as the integrand of the Lagrangian $L$.

\paragraph{Variational formulation in Lagrangian description.} In absence of irreversible processes and in absence of heat or matter exchange with the exterior, the entropy is conserved. This means that in Lagrangian variables the entropy is time independent, $S(t, \mathbf{X} )=S_0( \mathbf{X} )$. In this case, the equations of motion follow from the \textit{Hamilton principle}:
\begin{equation}\label{HP_fluids} 
\delta \int_{t_1}^{t_2} \!\!\int_ \mathcal{D}\mathfrak{L} \, d \mathbf{X} \, dt=0,
\end{equation} 
where the critical condition is taken with respect to arbitrary variations $ \delta \varphi $ of the Lagrangian trajectory $\varphi$, with $ \delta \varphi (t_1)= \delta \varphi (t_2)=0$.

\medskip

We shall now extend the Hamilton principle \eqref{HP_fluids} in order to incorporate the irreversible processes of viscosity and heat conduction. Following \cite{GBYo2016b}, we consider the variational condition 
\begin{equation}\label{VPLagr} 
\delta \int_{t_1}^{t_2} \!\!\int_ \mathcal{D} \big( \mathfrak{L} + (S- \Sigma ) \dot \Gamma \, \big) d \mathbf{X} \, dt =0,
\end{equation} 
subject to the \textit{phenomenological constraint}
\begin{equation}\label{KC}
\frac{\partial \mathfrak{L} }{\partial S}\dot \Sigma = -\mathbf{P} ^{\rm fr}: \nabla\dot \varphi  + \mathbf{J} _S \cdot \nabla \dot\Gamma  
\end{equation}
and with respect to variations $ \delta \varphi $, $ \delta S$, $ \delta \Sigma $, $ \delta \Gamma $ subject to the \textit{variational constraint}
\begin{equation}\label{VC}
\frac{\partial \mathfrak{L}}{\partial S}\delta \Sigma= - \mathbf{P} ^{\rm fr}: \nabla \delta \varphi  + \mathbf{J} _S \cdot \nabla \delta\Gamma
\end{equation} 
and with $\delta \varphi (t_i)=0$, $\delta\Gamma (t_i)=0$, $i=1,2$.

This variational formulation introduces two new functions, $\Sigma$ and $ \Gamma $, whose time derivative will be ultimately identified with the entropy generation rate density and the temperature, respectively. The tensor $\mathbf{P} ^{\rm fr}$ is the Piola-Kirchhoff viscous stress tensor and $ \mathbf{J} _S$ is the entropy flux density in Lagrangian representation. \textcolor{black}{The Piola-Kirchhoff viscous stress tensor is the Lagrangian object associated to the Eulerian viscous stress, see \cite{GBYo2016b}}. The link with the corresponding familiar Eulerian objects will be explained below. The notation ``$\,:\,$" indicates the contraction of tensors with respect to two indices.

The $ \delta $-notation in \eqref{VPLagr} indicates that we compute the variation of the functional with respect to all the fields, namely $\varphi , S, \Sigma , \Gamma $. The constraint \eqref{KC} is referred to as a \textit{phenomenological constraint}, since the expression of the thermodynamic fluxes $\mathbf{P} ^{\rm fr}$ and $ \mathbf{J} _S$ are obtained through phenomenological laws, also called parameterizations of the irreversible processes. It is a nonlinear constraint on the time derivatives of the field variables. We note that the right hand side of this constraint is of the generic form $\sum_ \alpha J_ \alpha \dot \Lambda  _\alpha $, for the thermodynamic fluxes $J _\alpha $ (here $ \mathbf{P} ^{\rm fr}$ and $ \mathbf{J} _S$) acting on the rate of thermodynamic displacements $\dot \Lambda _ \alpha  $ (here $ \dot \varphi $ and $ \dot \Gamma $), where the index $ \alpha $ refers to an irreversible process.
The constraint \eqref{VC} is referred to as a \textit{variational constraint}, since it is a constraint on the variations to be considered in \eqref{VPLagr}.
One passes from \eqref{KC} to \eqref{VC} by formally replacing each occurrence of a time derivative by a $ \delta $-variation, i.e., $\sum_ \alpha J_ \alpha \dot \Lambda  _\alpha \leadsto  \sum_ \alpha J_ \alpha  \delta \Lambda  _\alpha$. This is reminiscent of what happens in the Lagrange-d'Alembert principle in nonholonomic mechanics, see Remark \ref{rmk_NH} below. We refer to \cite{GBYo2016a,GBYo2016b} for several examples of phenomenological constraints in nonequilibrium thermodynamics of discrete and continuum systems.

\medskip

Taking the variations in \eqref{VPLagr}, using the variational constraint \eqref{VC} and collecting the terms proportional to $ \delta \varphi $, $ \delta \Gamma $, and $ \delta S$, we get the three conditions
\begin{align*}
&\frac{d}{dt}\frac{\partial \mathfrak{L} }{\partial \dot \varphi }+ \operatorname{DIV}  \left( \frac{\partial \mathfrak{L}}{\partial \nabla  \varphi } +\dot  \Gamma \frac{\partial \mathfrak{L}}{\partial S}^{-1}\mathbf{P} ^{\rm fr} \right) - \frac{\partial \mathfrak{L} }{\partial \varphi }=0,\\
&\dot  S=\operatorname{DIV} \left(\dot  \Gamma \frac{\partial \mathfrak{L} }{\partial S}^{-1} \mathbf{J} _S\right) + \dot \Sigma ,\qquad \qquad \dot\Gamma = - \frac{\partial \mathfrak{L}}{\partial S},
\end{align*} 
where $ \operatorname{DIV}$ denotes the divergence with respect to the labels $ \mathbf{X} $.
From the last equation, we have $ \dot \Gamma = - \frac{\partial \mathfrak{L} }{\partial S}=\mathfrak{T} $, the temperature in material representation. This attributes to $ \Gamma $ the meaning of \textit{\textcolor{black}{thermal} displacement}, as considered in \cite{GrNa1991} and introduced in \cite{He1884}. The second equation simplifies to $\dot S+ \operatorname{DIV} \mathbf{J} _S=\dot \Sigma$ and attributes to $ \dot \Sigma $ the meaning of \textit{entropy generation rate density}. From the first equation and the constraint, we thus get the system
\[
\left\{
\begin{array}{l}
\vspace{0.2cm}\displaystyle\frac{d}{dt}\frac{\partial \mathfrak{L} }{\partial \dot \varphi }+ \operatorname{DIV}  \left( \frac{\partial \mathfrak{L} }{\partial \nabla  \varphi } -\mathbf{P} ^{\rm fr} \right) - \frac{\partial \mathfrak{L} }{\partial \varphi }=0\\
\mathfrak{T} (\dot S + \operatorname{DIV} \mathbf{J} _S ) = \mathbf{P} ^{\rm fr}: \nabla\dot \varphi  - \mathbf{J} _S \cdot \nabla \mathfrak{T}  ,
\end{array}
\right.
\]
for the fields $ \varphi (t, \mathbf{X} )$ and $ S(t, \mathbf{X} )$. We will discuss the parameterization of the thermodynamic fluxes $ \mathbf{P} ^{\rm fr}$ and $ \mathbf{J} _S$ in terms of the thermodynamic forces below in the Eulerian description.

We leave to the reader the computation of the explicit form of these equations for the Lagrangian $ \mathfrak{L} $ of the dry atmosphere in \eqref{Lagrangian}. We shall only present the computation in Eulerian variables below.

\begin{remark}[Lagrange-d'Alembert principle in nonholonomic mechanics]\label{rmk_NH} {\rm We now comment on the analogy between our variational formulation and the Lagrange-d'Alembert principle used in nonholonomic mechanics. Let us consider a mechanical system with configuration variable $q$ and velocity $v= \dot q$. We assume that the system is subject to a linear constraint on velocity, i.e., a constraint of the form $ \omega (q) \cdot \dot q=0$, for 
a $q$-dependent linear map $\omega (q)$. Typical examples are rolling constraints. The extension of Hamilton's principle to nonholonomic systems (called the Lagrange-d'Alembert principle) consists in imposing that the action functional is critical with respect to variations $ \delta q$ subject to the linear constraint $ \omega (q) \cdot\delta q=0$. It formally follows by replacing the time derivative in the constraint $ \omega (q) \cdot \dot q=0$ by a $ \delta $-derivative, see, e.g., \cite{Bl2003}. In our case, the passage from \eqref{KC} to \eqref{VC} can be formally seen as a generalization of this approach, to the case of a nonlinear constraint. We refer to \cite{GBYo2016b} for an extensive discussion of the principle \eqref{VPLagr}--\eqref{VC}.
}
\end{remark} 

\color{black}
\begin{remark}[Variational formulation with Rayleigh dissipation function]\label{rmk_HR}{\rm A classical approach to include dissipation phenomena in Euler-Lagrange dynamics is due to \cite{Ra1877}. This approach applies when the work done by dissipative actions can be expressed in terms of a Rayleigh dissipation function $\mathcal{R}(q,\dot q)$ depending on the configuration $q$ and velocity $v=\dot q$ of the mechanical system. In abstract mechanical notations, the associated variational formulation takes the form
\begin{equation}\label{Rayleigh_Hamilton}
\delta \int _0^TL (q,\dot q)dt= \int_0^T \left\langle\frac{\partial \mathcal{R}}{\partial\dot q}, \delta q\right\rangle dt
\end{equation}
and yields the Euler-Lagrange equations with dissipative force
\begin{equation}\label{Rayleigh}
\frac{d}{dt}\frac{\partial L}{\partial \dot q}- \frac{\partial L}{\partial q}= - \frac{\partial \mathcal{R}}{\partial\dot q},
\end{equation}
where it is assumed $\big\langle \frac{\partial \mathcal{R}}{\partial\dot q}, \dot q\big\rangle\geq 0$. We refer, for instance, to \cite{dIMaSe2009} for an application of \eqref{Rayleigh_Hamilton} as a modelling tool in continuum mechanics.

When the Lagrangian only consists of a potential energy, $L(q,\dot q)=-U(q)$, then the Rayleighian defined as $\mathfrak{R}(q,\dot q)= \big\langle\frac{\partial U}{\partial q},\dot q\big\rangle + \mathcal{R}(q,\dot q)$ is considered. In this special case, the equations \eqref{Rayleigh} can be written as $\frac{\partial \mathfrak{R}}{\partial \dot q}=0$. This is sometimes called Rayleigh's principle of the least energy dissipation.

The principle \eqref{Rayleigh_Hamilton} has no relations with our variational formulation for thermodynamics, both in its nature and in the form of equations that it gives. In particular, the equations \eqref{Rayleigh} are dissipative, whereas the one deduced from our variational formulation are conservative for an isolate system in agreement with the first law of thermodynamics. In order to illustrate this discussion, we shall consider a finite dimensional thermodynamical system with only one scalar entropy variables $S$. In this case, given a Lagrangian $L(q,\dot q, S)$ and a friction force $F^{\rm fr}(q,\dot q, S)$, our variational formulation is (see \cite{GBYo2016a})
\begin{equation}\label{LdA_thermo} 
\delta \int_{ t _0 }^{ t _1 }L(q , \dot q , S)dt =0,
\end{equation}
where the curve $q(t)$ and $S(t)$ satisfy the constraint
\begin{equation}\label{Kinematic_Constraints} 
\frac{\partial L}{\partial S}(q, \dot q, S)\dot S  = \left\langle F^{\rm fr}(q, \dot q, S) , \dot q \right\rangle
\end{equation} 
and with respect to the variations $ \delta q $ and $\delta S$ subject to
\begin{equation}\label{Virtual_Constraints} 
\frac{\partial L}{\partial S}(q, \dot q, S)\delta S= \left\langle F^{\rm fr}(q , \dot q , S),\delta q \right\rangle .
\end{equation}
The principle \eqref{LdA_thermo}--\eqref{Virtual_Constraints} is a finite dimensional version of \eqref{VPLagr}--\eqref{VC} which, in addition, does not involve any heat conduction.
This principle gives the coupled equations of motion for the thermodynamical system as
\begin{equation}\label{SimpleSystem}
\frac{d}{dt}\frac{\partial L}{\partial \dot q}- \frac{\partial L}{\partial q}=F^{\rm fr}(q,\dot q,S), \qquad \frac{\partial L}{\partial S}\dot S=\left\langle F^{\rm fr}(q,\dot q,S), \dot q\right\rangle.
\end{equation}
In general, the system of equations in \eqref{SimpleSystem} cannot be recast in the form \eqref{Rayleigh} associated with the variational formulation \eqref{Rayleigh_Hamilton}. In addition, the force $F^{\rm fr}(q,\dot q,S)$ may not arise from a Rayleigh dissipation function.
}
\end{remark}

\begin{remark}[Local equilibrium hypothesis]{\rm
The variational formulation of nonequilibrium thermodynamic presented in this paper assumes the local equilibrium hypothesis, that is, the local and instantaneous relations between thermodynamic quantities in the nonequilibrium system are the same as for the system in equilibrium. This setting is sometimes referred to as classical irreversible thermodynamics. The relation between the thermodynamic fluxes and forces does not need to be linear in our variational formulation. Hence it is not restricted to linear irreversible thermodynamics.
}
\end{remark}
\color{black}

\paragraph{Variational formulation in Eulerian description.} In terms of Eulerian fields, the Lagrangian \eqref{Lagrangian} reads
\[
\ell( \mathbf{v} , \rho , s)= \int_ \mathcal{D}  \mathcal{L}( \mathbf{v} , \rho , s) \,d \mathbf{x},
\]
for the Lagrangian density $ \mathcal{L} $ given by
\begin{equation}\label{Lagr_dry_atm} 
\mathcal{L}( \mathbf{v} , \rho , s) = \rho\left( \frac{1}{2}| \mathbf{v} | ^2 + \mathbf{R}  \cdot \mathbf{v} - \Phi - u \left(s/ \rho , 1/ \rho \right)  \right) .
\end{equation} 
The Eulerian quantities $\gamma (t, \mathbf{x} )$ and $ \sigma (t, \mathbf{x} )$ associated to $ \Gamma (t, \mathbf{X} )$ and $ \Sigma (t, \mathbf{X} )$ are defined as 
\begin{equation}\label{def_Gamma_Sigma} 
\Gamma (t, \mathbf{X} )= \gamma (t, \varphi (t, \mathbf{X} )) \quad\text{and}\quad  \Sigma (t, \mathbf{X} )=  \sigma (t, \varphi (t, \mathbf{X} ))| \nabla  \varphi (t, \mathbf{X} )|;
\end{equation} 
the entropy flux density $ \mathbf{j} _s$ in Eulerian representation is related to $ \mathbf{J} _S$ as
\begin{equation}\label{def_JS} 
\nabla\varphi ( t,\mathbf{X} ) \cdot \mathbf{J} _S( t,\mathbf{X} )= | \nabla\varphi (t, \mathbf{X} )| \mathbf{j} _s(t,\varphi (t, \mathbf{X} ));
\end{equation} 
finally, the viscous stress $ \boldsymbol{\sigma} ^{\rm fr}$ 
is related to the Piola-Kirchhoff viscous stress $ \mathbf{P} ^{\rm fr}$ as
\begin{equation}\label{def_P} 
\mathbf{P} ^{\rm fr} (t, \mathbf{X} )\cdot \nabla\varphi (t, \mathbf{X} )= |\nabla\varphi (t, \mathbf{X} )| \boldsymbol{\sigma} ^{\rm fr}(t, \mathbf{X} ).
\end{equation}

By using these relations, we can rewrite the variational formalism \eqref{VPLagr}--\eqref{VC} entirely in terms of Eulerian variables as follows:
\begin{equation}\label{VP_Eulerian}
\delta  \int_{t_1}^{t_2} \!\! \int_ \mathcal{D} ( \mathcal{L} + (s- \sigma ) D_t \gamma ) \,d \mathbf{x}\, dt=0,
\end{equation} 
subject to the \textit{phenomenological constraint}
\begin{equation}\label{KC_Eulerian}
\frac{\partial \mathcal{L} }{\partial s}\bar D_t \sigma = - \boldsymbol{\sigma}  ^{\rm fr}: \nabla \mathbf{v}   + \mathbf{j} _s \cdot \nabla D_t\gamma   \\ 
\end{equation} 
and with respect to variations
\begin{equation}\label{variations} 
\delta \mathbf{v} =\partial _t \boldsymbol{\zeta} + \mathbf{v} \cdot \nabla \boldsymbol{\zeta} - \boldsymbol{\zeta}\cdot \nabla \mathbf{v} ,\;\;\;\; \delta \rho =- \operatorname{div}( \rho \boldsymbol{\zeta} ) , \;\;\;\;\delta s,\;\;\;\; \delta \sigma , \;\;\;\; \text{and}\;\;\;\;\delta \gamma,
\end{equation} 
that are subject to the \textit{variational constraint}
\begin{equation} \label{VC_Eulerian} 
\frac{\partial \mathcal{L} }{\partial s}\bar D_ \delta  \sigma = - \boldsymbol{\sigma}  ^{\rm fr}: \nabla \boldsymbol{\zeta}   + \mathbf{j} _s \cdot \nabla D_\delta \gamma 
\end{equation} with $ \boldsymbol{\zeta} (t_i)=0$ and  $ \delta \gamma (t _i )=0$, $i=1,2$.

The first two expressions in \eqref{variations} are obtained by taking the variations with respect to $\varphi $, $ \mathbf{u} $, and $ \rho $, of the relations $\dot \varphi (t, \mathbf{X} )= \mathbf{u} (t,\varphi (t, \mathbf{X} ))$ and $ \varrho  _0(\mathbf{X} )=  \rho (t, \varphi (t, \mathbf{X} ))| \nabla  \varphi (t, \mathbf{X} )|$, respectively, and defining $\boldsymbol{\zeta}(t, \mathbf{x} ) $ as $ \delta  \varphi (t, \mathbf{X} )=\boldsymbol{\zeta}  (t,\varphi (t, \mathbf{X} ))$. These formulas can be also directly justified by employing the Euler-Poincar\'e reduction theory on Lie groups, see \cite{HoMaRa2002}.

In \eqref{KC_Eulerian} and \eqref{VC_Eulerian}, we introduced the notations $D_tf:= \partial _t f+ \mathbf{v}\cdot \nabla f$, $\bar D_t f := \partial _t f + \operatorname{div}( f  \mathbf{v} )$, $D_ \delta f:= \delta  f+ \boldsymbol{\zeta} \cdot \nabla f$ and $\bar D_ \delta  f := \delta  f + \operatorname{div}( f  \boldsymbol{\zeta}  )$ for the Lagrangian time derivatives and variations of scalar fields and density fields.

\medskip

By applying \eqref{VP_Eulerian} and using the expression for the variations $ \delta \mathbf{v} $ and $ \delta \rho $, we find the condition
\begin{align*} 
&\int_{t_1}^{t_2} \!\!\int_ \mathcal{D} \left[  \left( \frac{\partial \mathcal{L} }{\partial \mathbf{v} }+( s- \sigma ) \nabla \gamma\right) \cdot (\partial _t \boldsymbol{\zeta} + \mathbf{v} \cdot \nabla \boldsymbol{\zeta} - \boldsymbol{\zeta} \cdot \nabla \mathbf{v} )-\frac{\partial \mathcal{L} }{\partial \rho } \operatorname{div}( \rho \boldsymbol{\zeta} ) \right.\\
&\left. \qquad\qquad\qquad \qquad\qquad \qquad + \left(\frac{\partial\mathcal{L} }{\partial s}+D_t\gamma  \right) \delta s  - \bar D_t( s- \sigma ) \delta \gamma -\delta \sigma D_t\gamma \right] d\mathbf{x}\, dt=0.
\end{align*} 
Using the variational constraint \eqref{VC_Eulerian} and collecting the terms proportional to $ \boldsymbol{\zeta} $, $ \delta s$, and $ \delta \gamma $, we obtain the three conditions
\begin{align*} 
&( \partial _t + \pounds _ \mathbf{v} ) \left( \frac{\partial \mathcal{L} }{\partial \mathbf{v} }+( s- \sigma ) \nabla \gamma\right) = \rho \nabla \frac{\partial \mathcal{L} }{\partial \rho }- \sigma \nabla D_t \gamma -\operatorname{div} \left( D_t \gamma\frac{\partial \mathcal{L} }{\partial s} ^{-1} \!\!\!\boldsymbol{\sigma}  ^{\rm fr} \right) \\
& \qquad\qquad \qquad \qquad\qquad \qquad\qquad\qquad\qquad \qquad \qquad + \operatorname{div} \left( D_t \gamma\frac{\partial \mathcal{L} }{\partial s} ^{-1} \!\!\!\mathbf{j} _s \right)\nabla\gamma, \\
& \bar D_t(s- \sigma )= \operatorname{div} \left( D_t \gamma\frac{\partial \mathcal{L} }{\partial s} ^{-1} \mathbf{j} _s \right), \qquad \qquad \frac{\partial \mathcal{L} }{\partial s}+ D_t\gamma =0,
\end{align*} 
where we introduced the Lie derivative notation $\pounds _ \mathbf{v} \mathbf{m}: = \mathbf{v} \cdot \nabla \mathbf{m} + \nabla  \mathbf{v} ^{\mathsf{T}} \cdot \mathbf{m} + \mathbf{m}\operatorname{div} \mathbf{v}  $ for a one-form density $ \mathbf{m}$ along a vector field $ \mathbf{v} $. Further computations finally yield the system
\begin{equation}\label{system_Eulerian} 
\left\{
\begin{array}{l}
\vspace{0.2cm}\displaystyle
( \partial _t + \pounds _ \mathbf{v} ) \frac{\partial \mathcal{L} }{\partial \mathbf{v} }= \rho \nabla \frac{\partial \mathcal{L} }{\partial \rho }+ s \nabla \frac{\partial \mathcal{L} }{\partial s }+ \operatorname{div} \boldsymbol{\sigma}  ^{\rm fr}\\
\vspace{0.2cm}\displaystyle  \frac{\partial \mathcal{L} }{\partial s}  (\bar D_t s + \operatorname{div} \mathbf{j} _s ) = -\boldsymbol{\sigma} ^{\rm fr} : \nabla \mathbf{v} - \mathbf{j} _s \cdot  \nabla \frac{\partial \mathcal{L} }{\partial s} \\
\bar D_t \rho =0,
\end{array}
\right.
\end{equation} 
whose last equation, the mass conservation equation, follows from the definition of $ \rho $ in terms of $\varrho _0 $.
These are the general equations of motion, in Lagrangian form, for fluid dynamics subject to the irreversible processes of viscosity and heat conduction. 
From one of the above conditions, we note that $ D_t \gamma = - \frac{\partial \mathcal{L} }{\partial s}= :T$, which attributes to $ \gamma $ the meaning of \textit{\textcolor{black}{thermal} displacement} in Eulerian variables. 
From the above conditions, we also note that the variable $ \sigma $ verifies
\begin{equation}\label{sigma_t} 
\bar D_t \sigma= \bar D_t s+ \operatorname{div} \mathbf{j} _s = \frac{1}{T} (\boldsymbol{\sigma} ^{\rm fr} : \nabla \mathbf{v} - \mathbf{j} _s \cdot  \nabla T).
\end{equation}
Therefore, $\bar D_t \sigma $ corresponds to the total \textit{entropy generation rate density} of the system. At this stage, the expressions of $\boldsymbol{\sigma} ^{\rm fr} $ and $ \mathbf{j} _s$ are still not specified.

In absence of irreversible processes (i.e., if $ \boldsymbol{\sigma} ^{\rm fr}=0$ and $ \mathbf{j} _s =0$), equations \eqref{system_Eulerian} recover the general form of the \textit{Euler-Poincar\'e equations} derived in \cite{HoMaRa2002} by Lagrangian reduction.

\medskip

We now write this system in the case of the Lagrangian $ \mathcal{L} $ of the dry atmosphere given in \eqref{Lagr_dry_atm}, but keeping a general expression for the internal energy. The partial derivatives are
\[
\frac{\partial  \mathcal{L}}{ \partial  \mathbf{v} }= \rho ( \mathbf{v} +\mathbf{R} ), \quad\frac{ \partial \mathcal{L}}{ \partial  \rho }= \frac{1}{2} | \mathbf{v} | ^2 +\mathbf{R} \cdot  \mathbf{v} - \Phi - u + \eta T-  v p, \quad\frac{\partial \mathcal{L} }{\partial s}= - \frac{\partial u}{\partial \eta}= -T.
\]
On one hand, we note that
\[
\rho \nabla \frac{\partial\mathcal{L} }{\partial \rho }+ s \nabla \frac{\partial\mathcal{L} }{\partial s }= \rho \nabla \left( \frac{1}{2} | \mathbf{v} | ^2  + \mathbf{R} \cdot\mathbf{v} \right)  - \rho \nabla \Phi -\nabla p,
\]
on the other hand, using $\bar D_t \rho =0$, we have
\[
( \partial _t + \pounds _ \mathbf{v} ) \left( \rho (\mathbf{v} +\mathbf{R} ) \right) =\rho ( \partial _t \mathbf{v} + \mathbf{v} \cdot \nabla (\mathbf{v}+  \mathbf{R})+ \nabla \mathbf{v} ^\mathsf{T} \cdot ( \mathbf{v} +\mathbf{R}  )).
\]
From \eqref{system_Eulerian} we thus get, as desired, the equations of motion for a heat conducting gas subject to the Coriolis, gravitational, and viscous forces:\begin{equation}\label{system_Eulerian_dry} 
\left\{
\begin{array}{l}
\vspace{0.2cm}\displaystyle
\rho ( \partial_t \mathbf{v} +\mathbf{v} \cdot \nabla \mathbf{v} +  2\boldsymbol{\Omega} \times \mathbf{v} )= - \rho \nabla \Phi- \nabla p +\operatorname{div} \boldsymbol{\sigma} ^{\rm fr}\\
\vspace{0.2cm}\displaystyle T (\bar D_t s + \operatorname{div} \mathbf{j} _s ) = \boldsymbol{\sigma} ^{\rm fr} : \nabla \mathbf{v} - \mathbf{j} _s \cdot  \nabla T\\
\bar D_t \rho =0.
\end{array}
\right.
\end{equation} 
The second equation in \eqref{system_Eulerian_dry} can be rewritten in terms of the temperature as
\begin{equation}\label{heat_equation} 
D_t T = - \rho c_s ^2 \Gamma  \operatorname{div} \mathbf{v}+ \frac{1}{\rho C_v} \big( \boldsymbol{\sigma} ^{\rm fr}:\nabla \mathbf{v}-\operatorname{div}(T\mathbf{j} _s)\big),
\end{equation} 
where $C_v (v,T)= T \frac{\partial \eta }{\partial T}(T,v)$ is the specific heat at constant volume,  $c_s ^2(v, \eta ) =\frac{\partial p}{\partial\rho } ( v, \eta )$ is the square of the speed of sound, and $\Gamma (p, \eta )= \frac{\partial T}{\partial p}(p, \eta )$ is the \textcolor{black}{adiabatic} temperature gradient (\textcolor{black}{or lapse rate}).

For meteorological applications, the potential temperature $ \theta $ is preferred as the temperature to describe the thermodynamic equation, since it turns out to be conserved in absence of viscosity and heat conduction. \textcolor{black}{The potential temperature is defined as the implicit solution of $\eta(T,p)=\eta(\theta,p_0)$, where $p_0$ is a given reference pressure. It can be written explicitly as}
\begin{equation}\label{def_theta} 
\theta (p,T):= T+\int_{p}^{p_0} \Gamma (p', \eta (p,T))dp'.
\end{equation} 
One obtains, from \eqref{heat_equation}, the potential temperature equation
\begin{equation}\label{theta_equation_general_dry} 
 D_t \theta =\frac{1}{ \rho C_p} \frac{\partial \theta }{\partial T} \big( \boldsymbol{\sigma} ^{\rm fr}: \nabla \mathbf{v} - \operatorname{div}(T \mathbf{j} _s)\big)=: \frac{\partial \theta }{\partial T}Q,
\end{equation} 
where $Q$ is the diabatic heating.

\paragraph{The case of dry air and ideal gases.} The above development is applicable to any state equation. For the case of a perfect gas, such as the dry air, there are several simplifications. The specific heat coefficient $C_v$ in \eqref{heat_equation} is a constant and from \eqref{state_equation}, we have $p=\rho R_dT$ in \eqref{system_Eulerian_dry} and $\rho c_s ^2 \Gamma =\frac{p}{ \rho C_v}$ in \eqref{heat_equation}, where $R_d$ is the gas constant for dry air. In this case, the potential temperature in \eqref{def_theta} recovers its usual expression $\theta = T/ \pi$, where $ \pi := (p/p_0)^{R_d/C_p}$ is the Exner pressure associated to $p_0$. One obtains, from \eqref{theta_equation_general_dry}, the potential temperature equation
\begin{equation}\label{theta_equation} 
 D_t \theta =\frac{\theta}{\rho T C_p} ( \boldsymbol{\sigma} ^{\rm fr}: \nabla \mathbf{v} - \operatorname{div}(T \mathbf{j} _s)).
\end{equation} 
It is the system \eqref{system_Eulerian_dry} with the second equation replaced by \eqref{theta_equation}, together with the state equation for the ideal gas $p=\rho R_dT$, that is traditionally used to describe the dynamics of a dry atmosphere, see, for instance, \cite{Gi1982}.
Our variational approach is however naturally expressed in terms of the entropy density rather than the potential temperature.

In some atmospheres, the perfect gas hypothesis may no longer be made. For example, for the atmosphere of Venus, while the state equation of an ideal gas can still be used, the specific heat depends on the temperature (\cite{Se1985}). An analytic approximation for this temperature dependence is given by $C_p(T)=C_{p_0} (T/T_0)^ \nu $, where $C_{p_0}, \nu , T_0$ are constants, see \cite{Le2010}, which yields a corresponding expression for the potential temperature, according to the general definition above. Of course, our variational formalism does apply to this case, as it does for any one component gas.

\paragraph{Heat exchanges.} So far, we have considered the atmosphere as an isolated system, i.e., with no exchange of work, heat or matter with its surroundings (space, ocean or earth's surface). The right hand side of the entropy equation in \eqref{system_Eulerian_dry} thus only consists of a net production of entropy due to the irreversible processes.

Heat exchanges between the atmosphere and its surroundings, such as radiative heating and cooling, surface sensible heat flux and surface latent heat flux, can be easily incorporated in our variational formulation, as long as these are considered as external processes.

We incorporate the heating into our variational formalism \eqref{VPLagr}--\eqref{VC} by modifying the phenomenological constraint \eqref{KC} to 
\begin{equation}\label{KC_heat}
\frac{\partial \mathfrak{L} }{\partial S}\dot \Sigma = -\mathbf{P} ^{\rm fr}: \nabla\dot \varphi  + \mathbf{J} _S \cdot \nabla \dot\Gamma  -R,
\end{equation}
where $R$ denotes the heating rate density in Lagrangian representation. The variational constraint \eqref{VC} is however kept unchanged. This follows a general principle on the variational formulation of thermodynamics stated in \cite{GBYo2016a,GBYo2016b}, namely, that external effects only affect the phenomenological constraint and not the variational constraint. Schematically, one passes from the phenomenological constraint to the variational constraint as $\sum_ \alpha J_ \alpha \dot \Lambda  _\alpha +P_{\rm ext} \leadsto  \sum_ \alpha J_ \alpha  \delta \Lambda _\alpha$, where $P_{\rm ext}$ denotes the power density associated to heat transfer between the system and the exterior.

In Eulerian representation, the variational formalism \eqref{VP_Eulerian}--\eqref{VC_Eulerian} is thus modified by adding the contribution of the heating in \eqref{KC_Eulerian} as
\begin{equation}\label{KC_Eulerian_radiation}
\frac{\partial \mathcal{L} }{\partial s}\bar D_t \sigma = - \boldsymbol{\sigma}  ^{\rm fr}: \nabla \mathbf{v}   + \mathbf{j} _s \cdot \nabla D_t\gamma -r,
\end{equation} 
where $R$ and $r$ are related as $R (t, \mathbf{X} )=  r (t, \varphi (t, \mathbf{X} ))| \nabla  \varphi (t, \mathbf{X} )|$. 
This results in a modification of the entropy equation in \eqref{system_Eulerian} into
\begin{equation}\label{VC_Eulerian_radiation}
T (\bar D_t s + \operatorname{div} \mathbf{j} _s ) = \boldsymbol{\sigma} ^{\rm fr} : \nabla \mathbf{v} - \mathbf{j} _s \cdot  \nabla T+ r
\end{equation} 
and corresponding modifications in the temperature and potential temperature equations.

\paragraph{Phenomenological relations and entropy production.} The phenomenological constraints \eqref{KC} and \eqref{KC_Eulerian} to be used in the variational formulations are determined by the expressions of the thermodynamic fluxes $ \boldsymbol{\sigma} ^{\rm fr}$ and $ \mathbf{j} _s$ in terms of the thermodynamic forces $ \nabla \mathbf{v} $ and $ \nabla T$. 
The phenomenological expression of these fluxes must be in agreement with the second law of thermodynamics, which requires that the internal entropy production, given in equation \eqref{sigma_t}, is positive. When only viscosity and heat conduction are considered, we have the well-known relations
\begin{equation}\label{friction_stress_NSF} 
\boldsymbol{\sigma} ^{\rm fr}=2 \mu  (\operatorname{Def} \mathbf{v})+ \left( \zeta - \frac{2}{3}\mu \right)(\operatorname{div} \mathbf{v} )  \delta \quad\text{and}\quad T\mathbf{j} _s= - k \nabla T \;\; \text{(Fourier law)},
\end{equation} 
with $\operatorname{Def} \mathbf{v}= \frac{1}{2} ( \nabla \mathbf{v} + \nabla \mathbf{v} ^\mathsf{T})$ the deformation tensor, and where $ \mu \geq 0 $ is the first coefficient of viscosity (shear viscosity), $ \zeta \geq 0$ is the second coefficient of viscosity (bulk viscosity), and $ k \geq 0$ is the thermal conductivity. In general, these coefficients may all depend on thermodynamic variables. For the atmosphere below 100 km, $\mu $ is so small ($ \mu =1.7 \times 10^{-5}\text{kg m}^{-1}\text{s}^{-1}$ for standard atmospheric conditions) that viscosity is negligible except in a thin layer within a few centimeters of the earth's surface where the vertical shear is very large. The second coefficient of viscosity is notoriously difficult to measure in compressible flows. A common assumption is Stokes' hypothesis $ \zeta =0$.

\section{Variational derivation of the thermodynamic of a moist atmosphere with rain process}\label{2_moist}

\color{black}
Moist air consists of dry air and water substance in its different phases, namely, water vapor, liquid water, and solid water. The gas component of moist air is thus a mixture of dry air and water vapor. The liquid and solid phases of water both consist of an airborne condensate (cloud)  and of a precipitating condensate (rain or snow). The associated mass densities are denoted as $\rho_d$ for dry air, $\rho_v$ for vapor, $\rho_c$ for airborne condensate, and $\rho_r$ for precipitating condensate. 
The total airborne water substance is $\rho_w=\rho_v+\rho_c$ and the total mass density of moist air is $\rho= \rho _d+  \rho _v+ \rho _c+\rho_r$.
The mass concentrations $q_k$ [kg kg$^{-1}$] and the molar concentrations $n_k$ [mol kg$^{-1}$] are defined as
\[
\rho _k=\rho q_k= \rho m_kn_k,\quad k=d,v,c,r,
\]
where $m_k$ [kg mol$^{-1}$] are the molecular weights. The vapor concentration $q_v$ is known as the specific humidity.

\medskip

The mass densities $\rho_v$ and $\rho_c$ are not independently predicted, but diagnostically separated from the predicted $\rho_w$, according to the saturation condition. More precisely, suppose that the air is in a state described by $(p,T,q_w,q_r)$ and consider the saturation specific humidity$^{2}$ $q^*(p,T)$. If the air is unsaturated, i.e. $q_w < q^*(p,T)$, then $q_v=q_w$ and $q_c=0$. If the air is saturated, i.e. $q_w\geq q^*(p,T)$, then $q_v=q_v^*(p,T,q_w,q_r)$ and $q_c=q_w-q_v$, where $q_v^*(p,T,q_w,q_r)$ is the specific humidity of saturated moist air$^{2}$.
\addtocounter{footnote}{1}
\footnotetext{The saturation specific humidity is given as $q^*(p,T)=\frac{\epsilon p^*(T)}{p-(1-\epsilon)p^*(T)}$, where $p^*(T)$ is the saturation vapor pressure, $ \epsilon = m_v/m_d$. In absence of rain, the specific humidity of saturated moist air is given as $q_v^*(p,T,q_w)= q_q\frac{ \epsilon p^*(T)}{p-p^*(T)}$, see, e.g., \cite{Sa2014}, for a derivation of these formulas.}The continuity equations for $\rho_d$, $\rho_w=\rho_v+\rho_c$, and $\rho_r$ have the general form, e.g. (3.1)--(3.3) in \cite{Oo2001},
\begin{equation}\label{continuity_equations}
\begin{aligned}
&\partial _t \rho _d+ \operatorname{div}(\rho _d \mathbf{v} + \mathbf{j} _d)=0\\
&\partial _t \rho _w+ \operatorname{div}(\rho _w \mathbf{v} + \mathbf{j} _w)=j_w\\
&\partial _t \rho _r+ \operatorname{div}(\rho _r (\mathbf{v} + \mathbf{v}_r^*)+\mathbf{j}_r)=j_r,
\end{aligned}
\end{equation}
where $j_w+j_r=0$ and $\mathbf{j}_d+\mathbf{j}_w+\mathbf{j}_r=0$, with $j_w$, $j_r$ the conversion rates, $\mathbf{j}_d$, $\mathbf{j}_w$, $\mathbf{j}_r$ the diffusion fluxes, and $\mathbf{v}^*_r$ is the terminal velocity of the precipitating component relative to the air.
These three equations are independent and may be combined into other convenient forms. In particular, the sum of the three gives the equations for the total mass density
\begin{equation}\label{total_mass}
\partial _t \rho + \operatorname{div}( \rho \mathbf{v}+\rho _r  \mathbf{v}_r^*)=0. 
\end{equation}

\medskip

\color{black}

The equation of state for the moist air can be expressed as
\begin{equation}\label{pressure_moist}
p= p_d+ p_v= (n_d+n_v) \rho R ^\ast T= (\rho _d R_d + \rho _vR_v)  T=\rho R_dT_v,
\end{equation} 
where $R_d= R ^\ast /m_d$ and $R_v= R ^\ast /m_v$ are the gas constant for dry air and water vapor, and $T_v=(q_d+ \epsilon ^{-1}  q_v)T$ is the virtual temperature, $ \epsilon = m_v/m_d$.
The specific internal energy of the moist air is given by
\begin{equation}\label{internal_energy_moist}
\begin{aligned} 
u&= q_d C_{vd}T + q_v(L(T)-R_vT)+ (q_v+q_c+q_r) C_{l}T\\
&=q_dC_{vd}T+  q_v(C_{vv}T+L_{00})+ (q_c+q_r) C_lT,
\end{aligned}
\end{equation} 
where
\begin{equation}\label{latent_heat} 
L(T)= L_v(T_0)+(C_{pv}-C_l)(T-T_0)= L_{00}+ (C_{pv}-C_{l})T
\end{equation} 
is the specific latent heat of vaporization, with $T_0$ a reference temperature, $C_{vk}$ and $C_{pk}$ are the specific heat capacities of dry air $(k=d)$ and water vapor $(k=v)$, and $C_{l}$ is the specific heat capacity of liquid water (see, e.g., Chapter IV in \cite{IrGo1981} for a derivation of this formula). \textcolor{black}{The internal energy \eqref{internal_energy_moist} is measured from that of liquid water at $0$ K}. In the Lagrangian formulation below it will be useful to consider the specific internal energy $u$ as a function of the variables $ \eta $, $v$, $n_d, n_v, n_c, n_r$, from which we have the thermodynamic relation $du= T d \eta -p dv+\sum_k \mu _k dn _k$, where $ \mu _k$, are the chemical potentials.

\color{black}

\medskip

We shall present the variational formulation in terms of the variables $\mathbf{v}, s, \rho_d, \rho_v, \rho_c, \rho_r$, considered as independent, from which we deduce an equation for $\rho_w=\rho_v+\rho_c$.
The saturation condition $q_v=q_v(p,T,q_w,q_r)$ and $q_c=q_c(p,T,q_w,q_r)$  is included afterwards in the resulting PDE. Inserting the saturation conditions directly in the Lagrangian would introduce discontinuities in its partial derivatives at the saturation point.

\medskip

\color{black}

We shall denote by $ \varrho _k (t, \mathbf{X} )=  \rho _k(t, \varphi (t, \mathbf{X} ))| \nabla  \varphi (t, \mathbf{X} )|$, $k=d,v,c,r$, the mass densities in Lagrangian representation and by $ \varrho = \varrho _d + \varrho _v + \varrho _c + \varrho _r$ the total mass density, which is constant by \eqref{total_mass}. The entropy density in Lagrangian representation is denoted as before by $ S(t, \mathbf{X} )$.

The Lagrangian of the moist atmosphere is given by
\begin{equation}\label{Lagrangian_atmosphere} 
\begin{aligned} 
&L( \varphi , \dot \varphi , \varrho _d , \varrho _v ,\varrho _c, S)\\
& \quad = \int_ \mathcal{D} \varrho  \left[ \frac{1}{2} \left| \dot\varphi\right| ^2 + \mathbf{R}  (\varphi ) \cdot \dot \varphi  -\Phi ( \varphi ) - u  \left(  \frac{S}{ \varrho } ,\frac{|  \nabla \varphi |}{ \varrho } , \frac{ \varrho _d }{m_d \varrho },\frac{ \varrho _v }{m_v \varrho },\frac{ \varrho _c }{m_c \varrho },\frac{ \varrho _r }{m_r \varrho } \right)\right] d \mathbf{X}\\
&\quad  =: \int_ \mathcal{D} \mathfrak{L}\, d \mathbf{X}.
\end{aligned}
\end{equation}
It has the same expression with the Lagrangian \eqref{Lagrangian} of the dry atmosphere, except for the last term, which is the internal energy of the moist air. An important difference is the role played by the mass densities. In \eqref{Lagrangian}, the mass density $ \varrho _0 $ is time independent and seen as a fixed parameter in the variational formulation. In \eqref{Lagrangian_atmosphere} the mass densities $ \varrho _k $, $k=d,v,c,r$ are time dependent variables, which will be fully involved in the variational formulation.

\subsection{Variational formulation in Lagrangian description}

\color{black}

We introduce the general notation
\[
\partial_t\rho_k+\operatorname{div}(\rho_k\mathbf{v}  +\mathbf{j}_k)+ \delta_{kr}\operatorname{div}\mathbf{j}^*_{s_r}=j_k,\qquad  k=d,v,c,r
\]
for the continuity equations, with diffusion fluxes $\mathbf{j}_k$ and conversion rates $j_k$ which verify $\sum_k \mathbf{j} _k=0$ and $\sum_kj _k =0$. The equation for the rain, $k=r$, has an additional term involving the flux $\mathbf{j}_r^*:= \rho_r\mathbf{v}^*_r$, with $\mathbf{v}^*_r$ the terminal velocity. We denote by $\mathbf{j}_{s_r}^*=s_r\mathbf{v}^*_r$ and $\boldsymbol{\sigma}_r^*=\rho\mathbf{v}\otimes \mathbf{v}^*_r$ the entropy flux and the stress associated to the rain process. In Lagrangian representation, these quantities will be denoted by $ \mathbf{J} _k $, $J_k$, $\mathbf{J}_r^*$, $\mathbf{J}_{S_r}^*$, and $\mathbf{P}^*_r$.

For the moist atmosphere with rain process, we propose the variational formalism 
\begin{equation}\label{VPLagr_atmosph} 
\begin{aligned}
&\delta   \int_0^T \!\! \int_ \mathcal{D}  \Big(  \mathfrak{L}  +\sum_k \varrho _k \dot W_k +  (S- \Sigma ) \dot \Gamma  \Big)  d \mathbf{X} \, dt \\
&\qquad\qquad\qquad+\underbrace{\int_0^T \int_{\mathcal{D}} \Big(\mathbf{P}^*_r:\nabla\delta\varphi+ \mathbf{J}_{S_r}^*\cdot \nabla \delta \Gamma+\mathbf{J}_r^*\cdot \nabla \delta W_r\Big) d \mathbf{X} \, dt}_{\text{Lagrangian virtual work of the rain process}}=0,
\end{aligned}
\end{equation} 
subject to the \textit{phenomenological constraint}
\begin{equation}\label{KC_atmosphere}
\frac{\partial \mathfrak{L} }{\partial S}\dot \Sigma= -\mathbf{P} ^{\rm fr}: \nabla\dot \varphi  + \mathbf{J} _S \cdot \nabla \dot\Gamma  + \sum_k( \mathbf{J} _k \cdot \nabla \dot W _k+  J_k \dot W _k )
\end{equation}  
and with respect to variations $ \delta \varphi $, $ \delta S$, $ \delta \Sigma $, $ \delta \Gamma $ subject to the \textit{variational constraint}
\begin{equation}\label{VC_atmosphere}
\frac{\partial \mathfrak{L}}{\partial S}\delta \Sigma = - \mathbf{P} ^{\rm fr}: \nabla \delta \varphi  + \mathbf{J} _S \cdot \nabla \delta\Gamma + \sum_k(\mathbf{J} _k \cdot \nabla \delta  W _k+  J_k \delta  W _k)
\end{equation} 
and with $\delta \varphi(t _i )= \delta\Gamma (t _i ) =\delta {W_k}(t_i )=0$, $i=1,2$.

\color{black}

The functions $ W_k$ in \eqref{VPLagr_atmosph}--\eqref{VC_atmosphere} will ultimately be identified with the \textit{thermodynamic displacements} associated to the irreversible processes (diffusion and phase transition) undergone by the substance $k$. This is in analogy with the \textcolor{black}{thermal} displacement $ \Gamma $ associated with the irreversible process of heat transfer that already appeared in the dry atmosphere earlier.

The $ \delta $-notation in \eqref{VPLagr_atmosph} indicates that we compute the variation of the functional with respect to all the field variables, namely, $\varphi , S, \Sigma , \Gamma , W_k, \varrho _k $. In a similar way with the case of the dry atmosphere, one passes from \eqref{KC_atmosphere} to \eqref{VC_atmosphere} by formally replacing each occurrence of a time derivative by a $ \delta $-variation, i.e., formally following the rule $\sum_ \alpha J_ \alpha \dot \Lambda  _\alpha \leadsto  \sum_ \alpha J_ \alpha  \delta \Lambda  _\alpha$.
The second term in \eqref{VPLagr_atmosph} represents the virtual work done on the system by the rain process.

Taking the variations in \eqref{VPLagr_atmosph}, using the virtual constraint \eqref{VC_atmosphere} and collecting the terms proportional to $ \delta \varphi $, $ \delta \Gamma $, $ \delta S$, $ \delta W _k  $, and $ \delta \varrho _k $, we get
\begin{align*}
&\frac{d}{dt}\frac{\partial \mathfrak{L} }{\partial \dot \varphi }+ \operatorname{DIV}  \left( \frac{\partial \mathfrak{L}}{\partial \nabla  \varphi } +\dot  \Gamma \frac{\partial \mathfrak{L}}{\partial S}^{-1}\mathbf{P} ^{\rm fr} +\mathbf{P}^*_r\right) - \frac{\partial \mathfrak{L} }{\partial \varphi }=0\\
&\dot  S=\operatorname{DIV} \left(\dot  \Gamma \frac{\partial \mathfrak{L} }{\partial S}^{-1} \mathbf{J} _S\right) + \dot \Sigma -\operatorname{DIV}\mathbf{J}_{S_r}^*, \qquad\qquad\qquad\qquad \quad \;\;\dot\Gamma = - \frac{\partial \mathfrak{L}}{\partial S}\\
&\dot\varrho_k - \operatorname{DIV} \left( \dot  \Gamma \frac{\partial \mathfrak{L} }{\partial S}^{-1} \mathbf{J}_k \right) +\delta_{kr}\operatorname{DIV}\mathbf{J}_r^* + \dot  \Gamma \frac{\partial \mathfrak{L} }{\partial S}^{-1} J_k=0 ,\qquad \qquad  \dot W _k =- \frac{\partial \mathfrak{L} }{\partial \varrho_k }. 
\end{align*} 
From the third and last equations, we have $ \dot \Gamma = - \frac{\partial \mathfrak{L} }{\partial S}=\mathfrak{T} $, the temperature in material representation, and $\dot W _k =- \frac{\partial \mathfrak{L} }{\partial \varrho_k }= \Upsilon _k $, a generalization of the chemical potential of substance $k$ in Lagrangian representation. The second equation thus reads $\dot S+ \operatorname{DIV} \mathbf{J} _S+\operatorname{DIV} \mathbf{J} _{S_r}^*=\dot \Sigma$ and attributes to $ \Sigma $ the same meaning as before. From the first and fourth equation and the constraint, we get the system
\begin{equation}\label{moist_material} 
\left\{
\begin{array}{l}
\vspace{0.2cm}\displaystyle\frac{d}{dt}\frac{\partial \mathfrak{L} }{\partial \dot \varphi }+ \operatorname{DIV}  \left( \frac{\partial \mathfrak{L} }{\partial \nabla  \varphi } -\mathbf{P} ^{\rm fr} +\mathbf{P}^*_r\right) - \frac{\partial \mathfrak{L} }{\partial \varphi }=0\\
\vspace{0.2cm}\displaystyle\dot \varrho _k + \operatorname{DIV} \mathbf{J} _k+\delta_{kr}\operatorname{DIV} \mathbf{J} _r^* = J_k ,\quad k=d,v,c,r \\
\mathfrak{T} (\dot S + \operatorname{DIV} \mathbf{J} _S +\operatorname{DIV} \mathbf{J} _{S_r}^*) = \mathbf{P} ^{\rm fr}: \nabla\dot \varphi  - \mathbf{J} _S \cdot \nabla \mathfrak{T}  - \sum_k(\mathbf{J} _k \cdot \nabla \Upsilon _k + J_k \Upsilon _k ),
\end{array}
\right.
\end{equation} 
for the fields $ \varphi (t, \mathbf{X} )$, $ \varrho _k (t, \mathbf{X} )$, and $ S(t, \mathbf{X} )$.
The parametrization of the thermodynamic fluxes $ \mathbf{P} ^{\rm fr}$, $ \mathbf{J} _S$, $ \mathbf{J} _k $, and $ J _k $  in term of the thermodynamic forces will be discussed in the Eulerian description below.

\color{black}
\begin{remark}[Variational formulation for multicomponent fluids]{\rm Note that in absence of the irreversible and rain processes, the constraints in the variational formulation \eqref{VPLagr_atmosph}--\eqref{VC_atmosphere} disappear and we recover Hamilton's principle for a multicomponent fluid. Hamilton's principles for multicomponent fluids have been considered earlier in the literature, e.g., \cite{BeDr1978}, \cite{Go1990}. In the latter paper it is applied to a Lagrangian representation associated with a reference space for each component $k$. In particular, the inverse maps $\psi_k=\varphi_k^{-1}$ are used as the independent fields in the Hamilton principle, where $\varphi_k$ is the Lagrangian field of the $k$-th component. The equations in terms of the barycentric motion are then deduced from the sum of the balance of momenta, energy, and entropy for each component, whereas in our case we directly obtain the equations in terms of the barycentric motion. Both the case in which the specific entropy of each component is constant and the case in which only the whole entropy of the multicomponent fluid is conserved are treated. The latter condition is imposed as a holonomic constraint in the Hamilton principle. This approach however does not include the irreversible processes from a variational perspective.}
\end{remark}
\color{black}

\subsection{Variational formulation in Eulerian description}

In terms of Eulerian fields, the Lagrangian \eqref{Lagrangian_atmosphere} becomes
\[
\ell( \mathbf{v} , \rho _d, \rho _v,\rho_c , \rho_r , s)= \int_ \mathcal{D}  \mathcal{L}(\mathbf{v} , \rho _d, \rho _v,\rho_c , \rho_r  , s) \,d \mathbf{x},
\]
for the Lagrangian density
\begin{equation}\label{l_moist} 
\mathcal{L}(\mathbf{v} , \rho _d, \rho _v,\rho_c , \rho_r  , s) = \rho\left( \frac{1}{2}| \mathbf{v} | ^2 + \mathbf{R}  \cdot \mathbf{v} - \Phi - u \Big( \frac{s}{ \rho }   ,  \frac{1}{\rho} ,\frac{\rho _d}{m_d \rho} ,  \frac{\rho _v}{m_v\rho }, \frac{\rho _c}{m_c\rho} , \frac{\rho _r}{m_r\rho }, \Big)  \right).
\end{equation} 
The Eulerian quantities associated to $ \Gamma (t, \mathbf{X} )$, $ \Sigma (t, \mathbf{X} )$,  $ \mathbf{J} _S(t, \mathbf{X} )$, $ \mathbf{P} ^{\rm fr}$, and $\mathbf{P}^*_r$ are defined as in \eqref{def_Gamma_Sigma}, \eqref{def_JS}, \eqref{def_P}. The diffusion flux densities $ \mathbf{j} _k $, $\mathbf{j}_r^*$, $\mathbf{j}^*_{s_r}$, the conversion rates densities $ j _k $, and the thermodynamic displacements $w _k $, $k=d,v,c,r$ are related to their Lagrangian counterpart as follows
\[
\nabla\varphi ( t,\mathbf{X} ) \cdot \mathbf{J} _k( t,\mathbf{X} )= | \nabla\varphi (t, \mathbf{X} )| \mathbf{j} _k(t,\varphi (t, \mathbf{X} )), \quad J_k (t, \mathbf{X} )= j_k (t, \varphi (t, \mathbf{X} )) | \nabla\varphi (t, \mathbf{X} )|
\]
and
\[
W_k (t, \mathbf{X} )= w_k (t, \varphi (t, \mathbf{X} )).
\]

\color{black}
With these definitions, the variational formulation \eqref{VPLagr_atmosph} in Eulerian variables reads
\begin{equation}\label{VP_Eulerian_atmosph}
\begin{aligned}
&\delta  \int_0^T \!\!\int_ \mathcal{D}\big( \mathcal{L} + \sum_k\rho _k D_t w _k+ (s- \sigma ) D_t \gamma \big)  d \mathbf{x}\, dt\\
&\qquad\qquad\qquad+\underbrace{\int_0^T \int_{\mathcal{D}} \Big(\boldsymbol{\sigma}^*_r:\nabla\boldsymbol{\zeta}+\mathbf{j}_{s_r}^*\cdot \nabla D_\delta \gamma+\mathbf{j}_r^*\cdot \nabla D_\delta w_r\Big) d \mathbf{x} \, dt}_{\text{Eulrian virtual work of the rain process}}=0,
\end{aligned}
\end{equation} 
subject to the \textit{phenomenological constraints}
\begin{equation}\label{KC_Eulerian_atmosph}
\frac{\partial \mathcal{L} }{\partial s}\bar D_t \sigma = - \boldsymbol{\sigma}  ^{\rm fr}: \nabla \mathbf{v}   + \mathbf{j} _s \cdot \nabla D_t\gamma +\sum_k(\mathbf{j} _k \cdot \nabla D_t w _k + j_k D_t w_k)
\end{equation} 
and with respect to variations $ \delta \mathbf{v} =\partial _t \boldsymbol{\zeta} + \mathbf{v} \cdot \nabla \boldsymbol{\zeta} - \boldsymbol{\zeta}\cdot \nabla \mathbf{v} $, $ \delta \rho _k$, $\delta w _k $, $\delta s$, $ \delta \sigma $, and $ \delta \gamma $ such that $ \boldsymbol{\zeta} $, $ \delta \sigma $ and $ \delta \gamma $ satisfy the \textit{variational constraint}
\begin{equation}\label{VC_Eulerian_atmosph}
 \frac{\partial \mathcal{L} }{\partial s}\bar D_ \delta  \sigma= - \boldsymbol{\sigma}  ^{\rm fr}: \nabla \boldsymbol{\zeta} +\mathbf{j} _s \cdot \nabla D_\delta \gamma  +\sum_k (\mathbf{j} _k \cdot \nabla D_ \delta  w _k + j_k D_ \delta  w_k )
\end{equation} 
with $ \delta w _k $, $ \delta \gamma $, and $\boldsymbol{\zeta} $ vanishing at $t=0,T$.
One notes the same rule as before when passing from the phenomenological constraint \eqref{KC_Eulerian_atmosph} to the variational constraint \eqref{VC_Eulerian_atmosph}.

By applying \eqref{VP_Eulerian_atmosph} and using the expression for the variations $ \delta \mathbf{v} $ and $ \delta \rho $, we find
{\fontsize{8.5pt}{9pt}\selectfont\begin{align*} 
&\int_0^T\!\!\int_ \mathcal{D} \left[\Big( \frac{\partial \mathcal{L} }{\partial \mathbf{v} }+ \sum_k\rho _k \nabla w_k +( s- \sigma ) \nabla \gamma\Big) \cdot (\partial _t \boldsymbol{\zeta} + \mathbf{v} \cdot \nabla \boldsymbol{\zeta} - \boldsymbol{\zeta} \cdot \nabla \mathbf{v} ) +\sum_k  \left(\frac{\partial \mathcal{L} }{\partial \rho _k} +D_t w _k \right) \delta \rho _k\right.\\
&  \left. - \sum_k  \bar D _t \rho _k \delta w _k + \left(\frac{\partial\mathcal{L} }{\partial s}+D_t\gamma  \right) \delta s  - \bar D_t( s- \sigma ) \delta \gamma -\delta \sigma D_t\gamma +\boldsymbol{\sigma}^*_r:\nabla\boldsymbol{\zeta}+ \mathbf{j}_{s_r}^*\cdot \nabla D_\delta \gamma+\mathbf{j}_r^*\cdot \nabla D_\delta w_r \right]d\mathbf{x} \, dt=0.
\end{align*}}
\!\!Using the variational constraint \eqref{VC_Eulerian_atmosph} and collecting the terms proportional to $ \boldsymbol{\zeta} $, $ \delta \gamma $, $ \delta s$, $ \delta w _k $, and $ \delta \rho _k$, we get
{\fontsize{9.5pt}{9pt}\begin{align*} 
&( \partial _t + \pounds _ \mathbf{v} ) \left( \frac{\partial \mathcal{L} }{\partial \mathbf{v} }+\sum_k \rho _k \nabla w _k +( s- \sigma ) \nabla \gamma \right) +\operatorname{div} \boldsymbol{\sigma}  ^*_r= - \sigma \nabla D_t \gamma -\operatorname{div} \left( D_t \gamma\frac{\partial \mathcal{L} }{\partial s} ^{-1} \!\!\!\boldsymbol{\sigma}  ^{\rm fr} \right) \\
& \qquad  \qquad   + \operatorname{div} \left( D_t \gamma\frac{\partial \mathcal{L} }{\partial s} ^{-1} \!\!\!\mathbf{j} _s \right)\nabla\gamma+ \sum_k \operatorname{div} \left( D_t \gamma\frac{\partial \mathcal{L} }{\partial s} ^{-1} \!\!\!\mathbf{j} _k \right) \nabla w _k -D_t \gamma\frac{\partial \mathcal{L} }{\partial s} ^{-1}\!\!\!j_k \nabla w _k, \\
&\qquad \qquad - \operatorname{div} \mathbf{j}_{s_r}^* \cdot \nabla \gamma - \operatorname{div} \mathbf{j}_r^* \cdot \nabla w_r\\ 
& \bar D_t(s- \sigma )= \operatorname{div} \left( D_t \gamma\frac{\partial \mathcal{L} }{\partial s} ^{-1} \mathbf{j} _s \right)-\operatorname{div} \mathbf{j}_{s_r}^* ,\qquad\qquad  \frac{\partial \mathcal{L} }{\partial s}+ D_t\gamma =0,\\
& \bar D_t\rho _k = \operatorname{div} \left( D_t \gamma\frac{\partial \mathcal{L} }{\partial s} ^{-1} \mathbf{j} _k \right) -\delta_{kr}\operatorname{div} \mathbf{j}_r^*-D_t \gamma\frac{\partial \mathcal{L} }{\partial s} ^{-1} j_k ,\qquad \qquad \frac{\partial \mathcal{L} }{\partial  \rho_k }+ D_tw_k =0.
\end{align*}}
\!\!Further computations yield the system
\begin{equation}\label{system_Eulerian_atmosphere_moist} 
\left\{
\begin{array}{l}
\vspace{0.2cm}\displaystyle
( \partial _t + \pounds _ \mathbf{v} ) \frac{\partial \mathcal{L} }{\partial \mathbf{v} } + \operatorname{div} \boldsymbol{\sigma}  ^*_r= \sum_{k=d,v,c}\rho _k \nabla \frac{\partial \mathcal{L} }{\partial \rho _k }+ s \nabla \frac{\partial \mathcal{L} }{\partial s }+ \operatorname{div} \boldsymbol{\sigma}  ^{\rm fr}\\
\vspace{0.2cm}\displaystyle\frac{\partial \mathcal{L} }{\partial s } (\bar D_t s + \operatorname{div} \mathbf{j} _s +\operatorname{div} \mathbf{j} _{s_r}^*) = - \boldsymbol{\sigma} ^{\rm fr} \!: \!\nabla \mathbf{v} - \mathbf{j} _s\! \cdot  \!\nabla \frac{\partial \mathcal{L} }{\partial s }- \sum_k\left(\mathbf{j} _k \!\cdot \! \nabla \frac{\partial \mathcal{L} }{\partial \rho _k }+j _k  \frac{\partial \mathcal{L} }{\partial\rho_k } \right) \\
\displaystyle \bar D_ t\rho _k + \operatorname{div} \mathbf{j} _k+\delta_{kr}\operatorname{div} \mathbf{j} _r^*=j _k,\quad k=d,v,c,r.
\end{array}
\right.
\end{equation} \color{black}
These are the general equations of motion for fluid dynamics with Lagrangian $\mathcal{L}$, subject to the irreversible processes of viscosity, heat conduction, diffusion, phase transition, \textcolor{black}{and with rain process}.
As before, $ \gamma $ has the meaning of the \textcolor{black}{thermal} displacement and $ \sigma $ is the entropy generation rate density, given here by
\begin{equation}\label{sigma_t_moist} 
\bar D_t \sigma = \frac{1}{T} \Big(\boldsymbol{\sigma} ^{\rm fr} : \nabla \mathbf{v} - \mathbf{j} _s \cdot  \nabla T+ \sum_{k=d,v,c}(\mathbf{j} _k \cdot  \nabla \frac{\partial \mathcal{L} }{\partial \rho _k }+j _k  \frac{\partial \mathcal{L} }{\partial\rho_k })\Big).
\end{equation}
The expressions of the thermodynamic fluxes $\boldsymbol{\sigma} ^{\rm fr} $, $ \mathbf{j} _s$, $ \mathbf{j} _k $, $ j_k $ will be reviewed later by using Onsager's reciprocal relations.

\medskip

For the Lagrangian $ \mathcal{L} $ of the moist atmosphere in \eqref{l_moist}, we have the partial derivatives
\[
\frac{\partial  \mathcal{L}}{ \partial  \mathbf{v} }= \rho ( \mathbf{v} +\mathbf{R} ), \quad\frac{ \partial \mathcal{L}}{ \partial  \rho_k }= \frac{1}{2} | \mathbf{v} | ^2 +\mathbf{R} \cdot  \mathbf{v} - \Phi -  \frac{ \mu _k }{m_k} , \quad\frac{\partial \mathcal{L} }{\partial s}= - \frac{\partial u}{\partial \eta}= -T,
\]
where the second expression is obtained by using the property $u= \frac{\partial u}{\partial v}v+ \frac{\partial u}{\partial \eta }\eta + \sum_k\frac{\partial u}{\partial n_k}n_k$ of the specific internal energy. Inserting these partial derivatives in \eqref{system_Eulerian_atmosphere_moist} and using $\sum_k \mathbf{j}_k  =0$, $\sum_k j_k  =0$, one gets, after several computations, the system\color{black}
\begin{equation}\label{system_Eulerian_moist} 
\left\{
\begin{array}{l}
\vspace{0.2cm}\displaystyle
\rho ( \partial_t \mathbf{v} +\mathbf{v} \cdot \nabla \mathbf{v} +  2\boldsymbol{\Omega} \times \mathbf{v} )+\bar D_t\rho (\mathbf{v}+\mathbf{R})+\operatorname{div} \boldsymbol{\sigma} ^*_r= - \rho \nabla \Phi- \nabla p +\operatorname{div} \boldsymbol{\sigma} ^{\rm fr}\\
\displaystyle T (\bar D_t s + \operatorname{div} \mathbf{j} _s+ \operatorname{div} \mathbf{j} _{s_r}^*) = \boldsymbol{\sigma} ^{\rm fr} : \nabla \mathbf{v} - \mathbf{j} _s \cdot  \nabla T - \sum_k(\mathbf{j} _k \cdot  \nabla \frac{\mu _k}{m_k} + j _k  \frac{\mu _k}{m_k})\\
 \bar D_ t\rho _k + \operatorname{div} \mathbf{j} _k+\delta_{kr}\operatorname{div} \mathbf{j} _r^*=j _k,\quad k=d,v,c,r.
\end{array}
\right.
\end{equation}
We note that in absence of the rain process, we have $\bar D_t \rho =0$, so that the last term of the left hand side of the balance of momentum vanishes. In presence of rain, we have $\bar D_t \rho =-\operatorname{div}\mathbf{j}_r^*$.

As we commented earlier, the variational principles requires the mass densities $\rho_k$ to be considered as independent variables. In practice, one deduces from them the continuity equation for the total airborne water substance $\rho_w=\rho_v+\rho_c$ and then obtains the values of $\rho_v$ and $\rho_c$ from the saturation conditions.
One cannot directly use $\rho_w$ as an independent in the variational principle since the internal energy depends explicitly both on $\rho_d$ and $\rho_c$.

\medskip

In terms of the temperature, the second equation in \eqref{system_Eulerian_moist} takes the form
\begin{equation}\label{heat_equation_moist} 
D_t T = - \rho c_s ^2 \Gamma  (\operatorname{div}\mathbf{u}+ v\operatorname{div}\mathbf{j}_r^* )+ \frac{1}{\rho C_v}\mathcal{Q}    + \sum_k\frac{\partial  T }{\partial q_k}D_tq_k,
\end{equation}
for $\mathcal{Q}$ and $D_tq_k$ given by
\begin{align*}
\mathcal{Q}&=\boldsymbol{\sigma} ^{\rm fr}:\nabla \mathbf{v}-\operatorname{div}(T\mathbf{j} _s) - \sum_k (\mathbf{j} _k \cdot \nabla \frac{\mu _k}{m_k} + j _k\frac{\mu _k}{m_k})+T(\eta\operatorname{div}\mathbf{j}_r^*-\operatorname{div}\mathbf{j}_{s_r}^*)\\
D_tq_k&= \frac{1}{\rho}(z _k- \operatorname{div}\mathbf{j}_k -(\delta_{kr}-q_k)\operatorname{div}\mathbf{j}_r^*),
\end{align*}
where $C_v= T \frac{\partial \eta }{\partial T}(v,T,q _d,q _v,q _c,q_r)$ is the specific heat of moist air at constant volume, $c_s ^2 =\frac{\partial p}{\partial\rho } ( v, \eta ,q _d,q _v,q _c,q_r)$ is the square of the speed of sound, 
$\Gamma = \frac{\partial T}{\partial p}(p, \eta ,q _d,q _v,q _c,q_r)$ is the \textcolor{black}{adiabatic} temperature gradient, and the partial derivative in the last term is taken for the temperature expressed as a function $T=T( v, \eta ,q _d,q _v,q _c,q_r)$. The pression equation reads
\begin{equation}\label{pression_equation_moist} 
\begin{aligned}
D_tp&=  - \rho c_s ^2(\operatorname{div}\mathbf{u}+ v\operatorname{div}\mathbf{j}_r^*) + \frac{\rho c_s ^2 \Gamma }{T} \mathcal{Q}  + \sum_k \frac{\partial  p}{\partial q_k}D_tq_k,
\end{aligned}
\end{equation}
where  the partial derivative in the last term is taken for the pression expressed as a function $p=p( v, \eta ,q _d,q _v,q _c,q_r)$.

For a multicomponent gas in meteorological applications, a generalisation of the potential temperature is defined as
\begin{equation}\label{def_theta_multi} 
\theta (p,T,q _d,q _v,q _c,q_r):= T+\int_{p}^{p_0} \Gamma (p', \eta (p,T,q _d,q _v,q _c,q_r),q _d,q _v,q _c,q_r)dp',
\end{equation} 
where $\Gamma$ is the \textcolor{black}{adiabatic} temperature gradient defined above and $p_0$ is a given a reference pressure. Using the following observations
\[
\frac{\partial \theta }{\partial p}=- \Gamma\frac{\partial \theta }{\partial T}, \qquad \;\; \frac{1}{\rho C_p}= \frac{1}{ \rho C_v}- \frac{\rho c_s ^2\Gamma ^2 }{T}, \qquad \;\; \left.\frac{\partial T}{\partial q_k}\right|_{p, \eta }=\left.\frac{\partial T}{\partial q_k}\right|_{v, \eta }-\Gamma\left .\frac{\partial p}{\partial q_k} \right|_{v, \eta },
\]
we obtains, from \eqref{heat_equation_moist} and \eqref{pression_equation_moist}, the potential temperature equation
\begin{align}
D_t \theta &= \frac{\partial\theta }{\partial T} \Big( \frac{1}{\rho C_p} \mathcal{Q} + \sum_k \left.\frac{\partial T}{\partial q_k} \right|_{p, \eta } D_tq_k  \Big)+\sum_k \frac{\partial \theta }{\partial q_k}D_tq_k \label{theta_equation_general} \\
&= \frac{1}{\rho C_p} \frac{\partial\theta }{\partial T} \mathcal{Q} + \sum_k \left.\frac{\partial\theta }{\partial q_k} \right|_{p, \eta } D_tq_k  ,\nonumber
\end{align}
where, as opposed to the equation \eqref{heat_equation_moist}, the partial derivative in \eqref{theta_equation_general}  is taken for the temperature expressed as a function $T=T(p, \eta ,q _d,q _v,q _c)$. For meteorological applications, it is advantageous to rewrite this equation by using the specific enthalpy. Defining the partial specific enthalpy and entropy $ h _k = \frac{\partial h}{\partial q _k }(p,T,  q _d,q _v,q _c)$, $ \eta  _k = \frac{\partial \eta }{\partial q _k }(p,T,  q _d,q _v,q _c)$ and noting the equalities $C_p\frac{\partial T}{\partial q_k}=-T \eta _k $ and $ \frac{\mu _k}{ m_k} = h _k - T \eta _k $, equation \eqref{theta_equation_general} becomes
\begin{equation}\label{theta_equation_general_enthapy} 
\begin{aligned}
D_t \theta  =& \frac{1}{ \rho C_p}  \frac{\partial\theta }{\partial T}\Big( \boldsymbol{\sigma} ^{\rm fr}: \nabla \mathbf{v} -\operatorname{div}(\mathbf{j} _s ^h ) - \sum_k (\mathbf{j} _k \cdot \nabla h _k  + j_kh _k) +T(\eta_r\operatorname{div}\mathbf{j}_r^*-\operatorname{div}\mathbf{j}_{s_r}^*)\Big) \\
&+  \sum_k \frac{\partial \theta }{\partial q_k}D_tq_k,
\end{aligned}
\end{equation} 
where $ \mathbf{j} _s ^h = T \big(\mathbf{j} _s -\sum_ k \eta _k\mathbf{j}_k \big)$ is the sensible heat flux.

\color{black}

\subsection{Potential vorticity and circulation theorem \textcolor{black}{with rain process}}

For the system \eqref{system_Eulerian_moist} with irreversible and rain processes, we consider the Rossby-Ertel potential vorticity $q$ defined as $ \rho q=\boldsymbol{\zeta}_a \cdot \nabla \theta$, where $ \boldsymbol{\zeta}_a = \operatorname{curl} \mathbf{v} + 2 \boldsymbol{\Omega}$ is the absolute vorticity and $ \theta $ is the potential temperature defined in \eqref{def_theta_multi}. A lengthy but standard computation yields the evolution equation of $q$ as
\begin{equation}\label{PV_moist} 
\rho D_t q= \operatorname{div} \left(  -\rho ^{-1} \nabla p \times \nabla \theta +\mathbf{X} \times\nabla \theta +   \boldsymbol{\zeta} _a\dot \theta \right)+ q\operatorname{div}\mathbf{j}_r^*,
\end{equation} 
where $ \dot \theta $ denotes the right hand side of \eqref{theta_equation_general}. 
Equation \eqref{PV_moist} follows from $\bar D_t\rho=- \operatorname{div}\mathbf{j}_r^*$ and from the evolution of the absolute vorticity $ \partial _t \boldsymbol{\zeta}_a + \operatorname{curl}( \boldsymbol{\zeta}_a\times \mathbf{v} )= \rho ^{-2} \nabla\rho \times \nabla p + \operatorname{curl}\mathbf{X}$ with $\mathbf{X}=\rho ^{-1} \operatorname{div} (\boldsymbol{\sigma} ^{\rm fr}-\boldsymbol{\sigma} ^*_r) + \rho ^{-1}\operatorname{div}\mathbf{j}_r^* ( \mathbf{v}+\mathbf{R}) $. Note that, unlike the case of a one-component gas, even in absence of irreversible processes, the potential vorticity is not materially conserved as it does not verify $D_tq=0$, this is due to the dependence of $p$ and $ \theta $ on the concentrations $q_d,q_v,q_c$, which makes the term $ \operatorname{div} (  -\rho ^{-1} \nabla p \times \nabla \theta)= \rho ^{-2} \nabla \rho \times \nabla p \cdot  \nabla \theta $ not zero in general.

We now consider the general Lagrangian system with irreversible \textcolor{black}{and rain} processes defined in  \eqref{system_Eulerian_atmosphere_moist}, associated to a given unspecified Lagrangian density $ \mathcal{L} $, and derive the corresponding potential vorticity evolution. In this general setting, we consider the potential vorticity $ \tilde{q} $ defined in terms of $ \mathcal{L} $ as $ \rho \tilde{q} =\boldsymbol{\zeta} _a \cdot \nabla \psi  $, where $ \boldsymbol{\zeta} _a = \operatorname{curl} \left(\rho ^{-1}  \partial \mathcal{L}/\partial \mathbf{v} \right)$ and $\psi$ is a scalar field satisfying an evolution equation $D_t \psi =\dot \psi$, for instance $\psi=\theta$ or $\psi=\eta$. The first equation in \eqref{system_Eulerian_atmosphere_moist} implies
\begin{equation}\label{zeta_a_multi}
\partial _t \boldsymbol{\zeta}_a + \operatorname{curl}( \boldsymbol{\zeta}_a\times \mathbf{v} )= \sum_ k\nabla q _k \times\nabla \frac{\partial  \mathcal{L} }{\partial \rho _k } +\nabla \eta  \times \nabla  \frac{\partial \mathcal{L} }{\partial s}  +\operatorname{curl}\mathbf{X}
\end{equation}
where $\mathbf{X}=\rho ^{-1} \operatorname{div} (\boldsymbol{\sigma} ^{\rm fr}-\boldsymbol{\sigma} ^*_r) - \rho ^{-2}\bar D_t\rho \frac{\partial\mathcal{L}}{\partial \mathbf{v}} $ and we recall that $\bar D_t\rho=-\operatorname{div}\mathbf{j}_r^*$.
From this equation and $D_ t \psi =\dot \psi $, one obtains the evolution of potential vorticity as
\begin{equation}\label{general_PV_moist}
\rho D_t  \tilde{q} = \operatorname{div} \Big(  \Big(   \sum_k  q  _k \nabla \frac{\partial \mathcal{L} }{\partial \rho _k }  + \eta \nabla \frac{\partial \mathcal{L} }{\partial s}  + \mathbf{X}\Big) \times\nabla \psi   + \dot \psi   \boldsymbol{\zeta} _a\Big) + \tilde{q}\operatorname{div}\mathbf{j}_r^*.
\end{equation} 
This general equation is useful for a unified treatment of potential vorticity for various approximations of the equations of atmospheric dynamics obtained via variational principles, in which the irreversible and rain processes are included, see \S\ref{Sec_PI}. This equation simplifies in absence of rain process, since in this case $\bar D_t\rho =0$ and $\boldsymbol{\sigma} ^*_r=0$.

%If, for example, $ \psi $ is chosen as the specific entropy $ \eta $, then \eqref{general_PV_moist} simplifies to
%\begin{equation}\label{general_PV_moist_s}
%\rho D_t  \tilde{q} = \operatorname{div} \Big( \Big(   \sum_k  q  _k \nabla \frac{\partial \mathcal{L} }{\partial \rho _k }   +  \rho ^{-1}  \operatorname{div} \boldsymbol{\sigma} ^{\rm fr} \Big) \times\nabla \eta    + \dot\eta    \boldsymbol{\zeta} _a\Big),
%\end{equation} 
%where, from the second equation in \eqref{system_Eulerian_atmosphere_moist}, $ \dot \eta $ is given by
%\begin{equation}\label{eta_multi} 
%\dot \eta =  \frac{1}{T\rho } \Big(  \boldsymbol{\sigma} ^{\rm fr}: \nabla \mathbf{v} - \operatorname{div}(T \mathbf{j} _s)-\sum_k ( \mathbf{j} _k %\cdot \nabla\frac{ \mu _k }{m_k }+ j _k\frac{\mu_k }{m_k}) \Big) , \quad T:= - \frac{\partial \mathcal{L} }{\partial s}.
%\end{equation} 

\medskip

Kelvin's circulation theorem for the system \eqref{system_Eulerian_moist} directly follows from the balance of momentum as
\[
\frac{d}{dt} \oint_{c_t} ( \mathbf{v} + \mathbf{R} ) \cdot d \mathbf{x} = \oint_{c _t } \rho ^{-1} \big(\operatorname{div} (\boldsymbol{\sigma}-\boldsymbol{\sigma}^*_r)  +\operatorname{div}\mathbf{j}_r^* (\mathbf{v}+\mathbf{R})\big)\cdot d \mathbf{x}  , \quad \boldsymbol{\sigma} =- p\delta + \boldsymbol{\sigma} ^{\rm fr},
\]
where $ c _t $ is a loop advected by the wind flow. Its more general version for system \eqref{system_Eulerian_atmosphere_moist} associated to a Lagrangian $ \mathcal{L} $ reads 
\[
\frac{d}{dt} \oint_{c_t} \frac{1}{ \rho } \frac{\partial \mathcal{L} }{\partial \mathbf{v} } \cdot d \mathbf{x} = \oint_{c _t } \left( \sum_k q _k  \nabla \frac{\partial\mathcal{L} }{\partial \rho_k }+   \eta \nabla \frac{\partial\mathcal{L} }{\partial s} + \mathbf{X}\right) \cdot d \mathbf{x}.
\]

\subsection{Phenomenological relations and entropy production}

The system of equations \eqref{system_Eulerian_moist} needs to be supplemented with phenomenological expressions for the \textit{thermodynamic fluxes} $J _\alpha $ (i.e., $ \boldsymbol{\sigma} ^{\rm fr}$, $ \mathbf{j} _s$, $ \mathbf{j} _k $, and $ j_k$) in terms of the \textit{thermodynamic forces} $X_\alpha $ (i.e., $ \operatorname{Def} \mathbf{v} $, $ \nabla  T$, $ \nabla  \frac{\mu _k }{m_k}  $, and $ \frac{\mu _k }{m_k}  $) compatible with the second law $I = J_ \alpha X _\alpha \geq 0$, where $I$ is the internal entropy production density which, in our case, takes the form
\begin{equation}\label{internal_entropy_production} 
I = \frac{1}{T} \Big( \boldsymbol{\sigma} ^{\rm fr} : \nabla \mathbf{v} - \mathbf{j} _s \cdot  \nabla T - \sum_k(\mathbf{j} _k \cdot  \nabla \frac{\mu _k}{m_k} + j _k  \frac{\mu _k}{m_k}) \Big) .
\end{equation} 
These phenomenological expressions determine the phenomenological constraint \eqref{KC_Eulerian_atmosph} and its associated variational constraint \eqref{VC_Eulerian_atmosph}  to be used in the variational formalism.

In order to rewrite the expression of entropy used in the variational derivation, in a form that is commonly used in meteorological applications, there are several steps that need to be undertaken. We shall describe them in details below. For simplicity, we do not consider the rain process in this section.

It is empirically accepted  that for a large class of irreversible processes and under a wide range of experimental conditions, the thermodynamic fluxes $ J_ \alpha $ are linear functions of the thermodynamic forces $ X^\alpha $, i.e., $J_ \alpha = \sum_ \beta \mathcal{L} _{ \alpha \beta } X_\beta $, where the transport coefficients $ \mathcal{L} _ { \alpha \beta }$ are state functions that must be determined by experiments or, if possible, derived by nonequillibrium statistical physics.

Besides defining a positive quadratic form, the coefficients $ \mathcal{L} _ { \alpha \beta }$ must also satisfy {\it Onsager-Casimir relations} (\cite{Onsager1931}, \cite{Ca1945}) due to the microscopic time reversibility and the {\it Curie principle} associated to material invariance (see, for instance, \cite{deGrootMazur1969}, \cite{KoPr1998}, \cite{Woods1975}). 
In the case of a multicomponent gas, decomposing the viscous stress tensor and deformation tensor in the sum of a traceless part and a diagonal part,
 $ \boldsymbol{\sigma} ^{\rm fr} =(\boldsymbol{\sigma} ^{\rm fr})^{(0)}+\frac{1}{3}( \operatorname{Tr}\boldsymbol{\sigma} ^{\rm fr} )\delta$ and  $ \operatorname{Def} \mathbf{v} = (\operatorname{Def} \mathbf{v}) ^{(0)}+\frac{1}{3}(\operatorname{div} \mathbf{v} )\delta $, we have the following phenomenological linear relations
\begin{equation}\label{Onsager}
-\begin{bmatrix}
\vspace{0.05cm}\mathbf{j} _s\\
\vspace{0.05cm}\mathbf{j} _d\\
\vspace{0.05cm}\mathbf{j} _v\\
\vspace{0.05cm}\mathbf{j} _c
\end{bmatrix}= 
\begin{bmatrix}
L _{ss} & L _{sd} & \cdots\\
L _{ds} & L _{dd} & \cdots\\
\vdots  & \vdots & \ddots
\end{bmatrix}
\begin{bmatrix}
\vspace{0.05cm}\nabla  T\\
\vspace{0.05cm}\nabla \frac{\mu_d}{m_d}\\
\vspace{0.05cm}\nabla \frac{\mu_v}{m_v}\\
\vspace{0.05cm}\nabla \frac{\mu_c}{m_c}
\end{bmatrix}, \qquad 
\begin{bmatrix}
\operatorname{Tr}\boldsymbol{\sigma} ^{\rm fr} \\
-j_d\\
-j_v\\
-j_c
\end{bmatrix}= 
\begin{bmatrix}
\mathcal{L} _{00} & \mathcal{L} _{0d}& \cdots\\
\mathcal{L} _{d0}&\mathcal{L} _{dd} &\cdots\\
\vdots & \vdots& \ddots
\end{bmatrix}
\begin{bmatrix}
\vspace{0.05cm}\frac{1}{3} \operatorname{div} \mathbf{v}  \\
\vspace{0.05cm}\frac{ \mu  _d}{m_d}\\
\vspace{0.05cm}\frac{ \mu  _v}{m_v}\\
\vspace{0.05cm}\frac{ \mu  _c}{m_c}
\end{bmatrix}
\end{equation}
and 
\[
(\boldsymbol{\sigma} ^{\rm fr}) ^{(0)}= 2 \mu (\operatorname{Def}\mathbf{v})^{(0)},
\]
where all the coefficients may depend on $(p, T , q_d, q _v , q _c )$. The first linear relation describes the vectorial phenomena of heat conduction (Fourier law), diffusion (Fick law) and their cross effects (Soret and Dufour effects). The second relation describes the scalar processes of bulk viscosity, phase changes, and their possible cross-phenomena. The third relation describes the tensorial process of shear viscosity.
The associated viscous stress reads
\[
\boldsymbol{\sigma} ^{\rm fr}= 2 \mu \operatorname{Def} \mathbf{v} + \left( \frac{1}{9} \mathcal{L} _{00} - \frac{2}{3} \mu \right) (\operatorname{div} \mathbf{v} ) \delta + \frac{1}{3} \sum_\ell\mathcal{L} _{0\ell}\frac{ \mu _\ell }{m _\ell }  \delta.
\]
The condition $\sum_k \mathbf{j} _k=0$ is satisfied if $\sum_kL_{ks}=\sum_kL _{k\ell}=0$, for all $\ell$. Similarly, condition $\sum_k j _k=0$ is satisfied if $\sum_k\mathcal{L} _{k0}=\sum_k\mathcal{L} _{k\ell}=0$, for all $\ell$.

The \textit{Onsager-Casimir} relations imply
\[
L_{sk}=L_{ks}, \qquad L_{kl}=L_{lk}, \qquad \mathcal{L} _{0k}=- \mathcal{L} _{k0},\qquad \mathcal{L} _{kl}= \mathcal{L} _{lk}, \qquad \text{for all $k,l=d,v,c$},
\]
see, e.g., \cite{deGrootMazur1969}.

We now explain how these general identities relate to those used in meteorological applications. In this case, the internal entropy production due to the vectorial processes in \eqref{internal_entropy_production} is usually written in terms of the sensible heat flux $ \mathbf{j} _s ^h = T \big(\mathbf{j} _s -\sum_ k \eta _k\mathbf{j}_k \big)$ and the thermodynamic forces $ \frac{ \nabla T}{T}$ and  $(\nabla \frac{\mu_k}{m_k})_T$ as
\begin{equation}\label{rewriting_vectorial} 
- \mathbf{j} _s^h \cdot \frac{1}{T}  \nabla T - \sum_{k=d,v,c}\mathbf{j} _k \cdot  \Big( \nabla \frac{\mu _k}{m_k} \Big) _T,
\end{equation} 
where $(\nabla \frac{\mu_k}{m_k})_T$ denotes the gradient of the function $\frac{\mu_k}{m_k}$ with respect to the variables $(p,q_d, q _v , q _c )$ only, the temperature being seen as a parameter.
The linear phenomenological relations are written in this case as
\[
-\begin{bmatrix}
\vspace{0.05cm}\mathbf{j} _s ^h \\
\vspace{0.05cm}\mathbf{j} _d\\
\vspace{0.05cm}\mathbf{j} _v\\
\vspace{0.05cm}\mathbf{j} _c\\
\end{bmatrix}= 
\begin{bmatrix}
A _{ss} &A _{sd}& \cdots\\
A _{ds}&A_{dd} &\cdots\\
\vdots & \vdots& \ddots
\end{bmatrix}
\begin{bmatrix}
\vspace{0.05cm}\frac{\nabla  T}{T}\\
\vspace{0.05cm}(\nabla \frac{\mu_d}{m_d})_T\\
\vspace{0.05cm}(\nabla \frac{\mu_v}{m_v})_T\\
\vspace{0.05cm}(\nabla \frac{\mu_c}{m_c})_T
\end{bmatrix}.
\]
One now observes that the matrices
\[
\mathsf{L}=\begin{bmatrix}
L _{ss} & L _{sd}& L_{sv}& L_{sc}\\
L _{ds} & L _{dd}& L_{dv}& L_{dc}\\
L _{vs} & L _{vd}& L_{vv}& L_{vc}\\
L _{cs} & L _{cd}& L_{cv}& L_{cc}
\end{bmatrix} \quad\text{and}\quad \mathsf{A}=\begin{bmatrix}
A _{ss} & A _{sd}& A_{sv}& A_{sc}\\
A _{ds} & A _{dd}& A_{dv}&A_{dc}\\
A _{vs} & A _{vd}& A_{vv}& A_{vc}\\
A _{cs} & A _{cd}& A_{cv}& A_{cc}
\end{bmatrix}
\]
are related as
\[
\mathsf{A}=\mathsf{M}\mathsf{L}\mathsf{M}^T, \quad \text{for} \quad \mathsf{M}= \begin{bmatrix}
T & -T \eta _d& -T \eta _v& -T \eta _c\\
0 &1 &0&0\\
0 &0 &1&0\\
0 &0 &0&1
\end{bmatrix}.
\]
Since $\mathsf{M}$ is invertible $(T>0)$, it follows that $\mathsf{L}$ is symmetric and positive if and only if $\mathsf{A}$ is symmetric and positive. So, one can equivalently  apply Onsager's reciprocal relation to the fluxes and forces $( \mathbf{j} _s ^h , \mathbf{j} _k )$ and $( \nabla T/T, (\nabla \mu_k /m_k )_T)$ or to the fluxes and forces $( \mathbf{j} _s  , \mathbf{j} _k )$ and $( \nabla T ,\nabla \mu_k /m_k )$. The former being the one used in meteorological applications, the latter being the one naturally associated to our variational formulation, see \eqref{KC_Eulerian_atmosph}, with
\[
\nabla T= \nabla D_t \gamma \quad\text{and}\quad \nabla \mu _k = \nabla D_t w _k,
\]
for the thermodynamic displacements $ \gamma $ and $ w_k $.

Furthermore, since $\sum_{k=d,v,c} \mathbf{j} _k =0$, it is possible to eliminate one of the flux, say $ \mathbf{j} _d $, and to rewrite the entropy production associated to the diffusion processes as $\sum_{k=v,c}\mathbf{j} _k \cdot  \nabla \big( \frac{ \mu _k }{m_k }-\frac{ \mu _d}{m_d} \big)$, similarly for $\sum_{k=v,c}   \mathbf{j} _k \cdot  \nabla \big( \frac{ \mu _k }{m_k }-\frac{ \mu _d}{m_d} \big)_T$. One then checks that imposing the Onsager reciprocal relations and positiveness of the matrix in this reduced form turns out to be equivalent to imposing them in the above form, where the conditions $\sum_kL _{ks}=\sum_kL _{k\ell}=0$ and $\sum_kA_{ks}=\sum_kA _{k\ell}=0$, for all $\ell$, are assumed.
It is this reduced form that we will use below for the moist atmosphere.

\subsection{Entropy production in the moist atmosphere}

So far we have not specified the state equation, so the above development is valid for any multicomponent gas. Let us now assume the relations \eqref{pressure_moist} and \eqref{internal_energy_moist}. For simplicity, we also assume that the continuity equations for $ \rho _k$, $k=d,v,c$ satisfy the conditions $ \mathbf{j} _c=0$, $j_d=0$, and hence $ \mathbf{j} _d+ \mathbf{j} _v=0$ and $j_v+j_c=0$ hold.

Definition \eqref{def_theta_multi} yields the expression of the potential temperature as
\[
\theta (T,p, q _d , q _v , q _c )= T \left( \frac{p _0 }{p}  \right) ^ {\frac{C_p-C_v}{C_p} }, 
\]
where $C_v=q_d C_{vd}+q_vC_{vv}+  q_cC_l$ and $C_p=q_d C_{pd}+q_vC_{pv}+  q_cC_l$. The potential temperature equation \eqref{theta_equation_general_enthapy} becomes
\[
\rho D_t \theta =\frac{1}{C_p}  \frac{ \theta }{T}\Big( \boldsymbol{\sigma} ^{\rm fr}: \nabla \mathbf{v} -\operatorname{div}(\mathbf{j} _s ^h ) - (C _{pv} -C_{pd})\mathbf{j} _v \cdot   \nabla T -L(T) j _v \Big)+ ( \theta_d -\theta _v ) \operatorname{div} \mathbf{j} _v + (\theta _v - \theta _c ) j_v  ,
\]
where $L(T)=h_v-h_c$ is the specific latent heat of vaporization defined in \eqref{latent_heat}.

The internal entropy production \eqref{internal_entropy_production}, with the vectorial processes rewritten using \eqref{rewriting_vectorial}, becomes  
\begin{equation}\label{internal_entropy_production_atmosph} 
I = \frac{1}{T} \Big(\boldsymbol{\sigma} ^{\rm fr} : \nabla \mathbf{v} - \mathbf{j} _s ^h \cdot  \frac{1}{T}\nabla T -  \mathbf{j} _v \cdot  \nabla \Big( \frac{\mu _v}{m_v}- \frac{ \mu _d }{m_d} \Big)_T   - j _v \Big( \frac{\mu _v}{m_v}- \frac{ \mu _c }{m_c}  \Big) \Big)\end{equation} 
so that the parameterization of the vectorial and scalar processes is of the form
{\fontsize{9.5pt}{10pt}\begin{equation}\label{param_flux} 
-\begin{bmatrix}
\vspace{0.1cm}\mathbf{j} _s ^h \\
\mathbf{j} _v
\end{bmatrix}
=
\begin{bmatrix}
\vspace{0.1cm}A _{ss} & A _{sv}\\
A _{vs}&A _{vv}
\end{bmatrix}
\begin{bmatrix}
\vspace{0.1cm}\nabla  T /T\\
\nabla \big(\frac{\mu_v}{m_v}-\frac{\mu_d}{m_d}\big)_T 
\end{bmatrix}, \quad 
\begin{bmatrix}
\vspace{0.1cm}\operatorname{Tr}\boldsymbol{\sigma} ^{\rm fr} \\
-j_v
\end{bmatrix}= 
\begin{bmatrix}
\vspace{0.1cm}\mathcal{L} _{00} & \mathcal{L} _{0v}\\
\mathcal{L} _{v0}&\mathcal{L} _{vv}
\end{bmatrix}\begin{bmatrix}
\vspace{0.1cm}\frac{1}{3} \operatorname{div} \mathbf{v}  \\
\frac{\mu_v}{m_v}-\frac{\mu_c}{m_c}
\end{bmatrix},
\end{equation}}
\!\!where the coefficients verify the Onsager-Casimir relations $A_{sv}=A_{vs}$ and $ \mathcal{L} _{0v}=- \mathcal{L} _{v0}$. In \eqref{param_flux}, the first phenomenological relation describes the processes of diffusion, heat conduction and thermo-diffusion. The matrix elements are related to the coefficients associated to these three processes.
The second relation in \eqref{param_flux} describes the coupling of viscous processes and phase changes.

In the case of moist air, equations \eqref{internal_entropy_production_atmosph} and  \eqref{param_flux} can be explicitly given in terms of the variables $(p,T,  q _d , q _v , q _c )$ as
\begin{equation}\label{explicit} 
\nabla \big(\frac{\mu_v}{m_v}-\frac{\mu_d}{m_d}\big)_T= T\Big( R_v\frac{ \nabla p_v}{p_v}- R_d \frac{ \nabla p _d }{p _d } \Big)\quad\text{and}\quad \frac{\mu _v}{m_v}- \frac{ \mu _c }{m_c}= R_v T \ln \frac{p_v}{p ^\ast (T)},
\end{equation} 
where $ p ^\ast (T)$ is the saturation vapor pressure$^{3}$. 
\addtocounter{footnote}{1}
\footnotetext{The saturation vapor pressure is $p ^\ast (T)= p _0^\ast \left( \frac{T}{T _0 } \right)  ^ {\frac{C_{pv}-C_l}{R_v}} \exp \left[\frac{L_{00}}{R_v}\left(\frac{1}{T_0}-\frac{1}{T} \right) \right]$, see, e.g., \cite{CuWe1999}.}

The expression of the internal entropy production \eqref{internal_entropy_production_atmosph}, with the relations \eqref{explicit}, is of fundamental use for the study of the entropy budget of the atmosphere, see, e.g., \cite{PaHe2002a,PaHe2002b}.

\medskip

The inclusion of the airborne ice component in our variational formulation is possible and does not present any supplementary difficulties.
Our variational formalism also allows the inclusion of chemical reactions in the dynamics. This can be achieved by combining the variational formulation presented earlier with the variational formulation for chemical reactions presented in \cite{GBYo2016b}.

%\todo{More details? where do these chemical reactions take place? presumably in the dry air.}

External heat sources and sinks, such as radiative exchange and heat exchange at the earth's surface, can be easily included in our variational formulation, in a similar way with the case of the dry atmosphere in \eqref{KC_heat}--\eqref{VC_Eulerian_radiation} earlier.

\color{black}

\section{Pseudoincompressible approximation}\label{Sec_PI}

In order to illustrate the efficiency of our variational formulation as a modeling tool in atmospheric thermodynamics, we derive a pseudoincompressible model for moist atmospheric thermodynamics with general equations of state and subject to the irreversible processes of viscosity, heat conduction, diffusion, and phase transition. For simplicity, we ignore the rain process in this Section, but it can be included exactly as earlier.

Soundproof models arise from the need to remove the fastest-moving atmospheric and oceanic waves, the sound waves, which can badly affect numerical simulation by forcing the desired low-frequency circulations with high-frequency oscillations. Frequently applied soundproof models are the Boussinesq approximation (\cite{Bo1903}), the anelastic approximation (\cite{OgPh1962}, \cite{LiHe1982}), and the pseudoincompressible approximation (\cite{Du1989}).
%We refer to \cite{Kl2009} for a study of the range of scales for which the anelastic and pseudoincompressible fluid approximations are valid using the method of asymptotic expansions.
The pseudoincompressible approximation has been initially derived for the ideal gas. 
Extension to general equations of state that preserve energy and potential vorticity was given in  \cite[\S4.1]{VaLeBrWoZw2013}.

The inclusion of irreversible processes in pseudoincompressible models, as needed for the description of moist pseudoincompressible atmospheric motion, is a delicate issue whose thermodynamic consistency must be appropriately ensured. Thermodynamically consistent moist pseudoincompressible models were presented in \cite{KlPa2012} and \cite{ONKl2014}.

In this Section we derive a thermodynamically consistent pseudoincompressible model for moist atmospheric thermodynamics with general equations of state and subject to the irreversible processes of viscosity, heat conduction, diffusion, and phase transition. We shall obtain the model by including the pseudoincompressible condition as a holonomic constraint in the variational formulation of nonequilibrium thermodynamics presented in \S\ref{2_moist}. In absence of thermodynamical effect, the variational formulation recovers the one developed in \cite[\S4.1]{VaLeBrWoZw2013}.

For the sake of brevity, we shall only present the variational formulation in the Eulerian description.
Also, we shall not give all the intermediate details and directly present the final equations. A more detailed description and thorough study of our model will be given in a future work.

\paragraph{Variational formulation.} 
Let us consider a hydrostatically balanced, stratified reference configuration with background density $ \rho _0( \mathbf{x} )$, pressure $p_0(\mathbf{x} )$, temperature $T_0( \mathbf{x} )$, and entropy $s_0( \mathbf{x} )$ with
\[
\nabla p_0=- \rho _0 \nabla \Phi .
\]
The pseudoincompressible constraint on the state variables is given by
\begin{equation}\label{PI_constraint} 
\mathcal{C} ( \rho _d,\rho _v,\rho _c  , s):= p(v,\eta,q _d,q_v,q _c )-p_0,
\end{equation}
so that $\mathcal{C}=0$ imposes the pressure to be equal to its reference value. It generalizes the constraint used in \cite{VaLeBrWoZw2013} to the multicomponent case.
From the Lagrangian density $ \mathcal{L} $ of the moist atmosphere and the constraint $ \mathcal{C} $, we define the density $ \mathcal{K} $:
\begin{equation}\label{K} 
\mathcal{K} ( \mathbf{v} ,  \rho _d,\rho _v,\rho _c, s, \lambda  )= \mathcal{L} ( \mathbf{v} ,  \rho _d,\rho _v,\rho _c, s)-\lambda \mathcal{C} ( \rho _d,\rho _v,\rho _c,s),
\end{equation} 
where we introduced a Lagrange multiplier $ \lambda (t,\mathbf{x} )$.

Our pseudoincompressible model for the thermodynamics of the moist atmosphere is obtained by applying the variational formalism \eqref{VP_Eulerian_atmosph}--\eqref{VC_Eulerian_atmosph} to the density \eqref{K}, i.e., we consider
\begin{equation}\label{VP_Eulerian_PI}
\delta  \int_0^T\!\!\int_ \mathcal{D}\Big[ \mathcal{K}+ \rho _d D_t w _d + \rho _v D_t w _v+\rho _c D_t w _c + (s- \sigma ) D_t \gamma \Big]  d \mathbf{x}\, dt=0,
\end{equation} 
subject to the \textit{phenomenological constraints}
\begin{equation}\label{KC_Eulerian_PI}
\frac{\partial \mathcal{K} }{\partial s}\bar D_t \sigma = - \boldsymbol{\sigma}  ^{\rm fr}: \nabla \mathbf{v}   + \mathbf{j} _s \cdot \nabla D_t\gamma +\sum_k\left( \mathbf{j} _k \cdot \nabla D_t w ^k + j_k D_t w^k\right)
\end{equation} 
and with respect to variations $ \delta \mathbf{v} =\partial _t \boldsymbol{\zeta} + \mathbf{v} \cdot \nabla \boldsymbol{\zeta} - \boldsymbol{\zeta}\cdot \nabla \mathbf{v} $, $ \delta \rho _k$, $\delta w _k $, $\delta s$, $ \delta \sigma $, $ \delta \gamma $, and $ \delta \lambda $ such that $ \boldsymbol{\zeta} $, $ \delta \sigma $ and $ \delta \gamma $ satisfy the \textit{variational constraint}
\begin{equation}\label{VC_Eulerian_PI}
\frac{\partial \mathcal{K} }{\partial s}\bar D_ \delta  \sigma= - \boldsymbol{\sigma}  ^{\rm fr}: \nabla \boldsymbol{\zeta} +\mathbf{j} _s \cdot \nabla D_ \delta \gamma +\sum_k \left( \mathbf{j} _k \cdot \nabla D_\delta  w ^k + j_k D_\delta  w^k\right)
\end{equation}
with $ \delta w _k $, $ \delta \gamma $, and $\boldsymbol{\zeta} $ vanishing at $t=0,T$.

\paragraph{Pseudoincompressible thermodynamics.} Taking the variation in \eqref{VP_Eulerian_PI}, using \eqref{KC_Eulerian_PI} and \eqref{VC_Eulerian_PI} and integrating by parts to isolate the free variations, we get the system
\begin{equation}\label{Lagrangian_form_Eulerian_PI} 
\left\{
\begin{array}{l}
\vspace{0.2cm}\displaystyle
\partial _t \frac{\partial \mathcal{L} }{\partial \mathbf{v} }  +\pounds _ \mathbf{v} \frac{\partial \mathcal{L} }{\partial \mathbf{v} } =\sum_{k=d,v,c} \rho _k \nabla  \frac{\partial \mathcal{K} }{\partial \rho_k }  + s \nabla  \frac{\partial \mathcal{K} }{\partial s}  +\operatorname{div} \boldsymbol{\sigma} ^{\rm fr}\\
\displaystyle\vspace{0.2cm} \frac{\partial \mathcal{K} }{\partial s } ( \bar D_t s +\operatorname{div} \mathbf{j} _s)= - \boldsymbol{\sigma} ^{\rm fr}: \nabla \mathbf{v}  - \mathbf{j} _s \cdot\nabla   \frac{\partial \mathcal{K} }{\partial s}-\sum_k \Big( \mathbf{j} _k\cdot \nabla   \frac{\partial \mathcal{K} }{\partial \rho_k }+  j_k   \frac{\partial \mathcal{K} }{\partial \rho_k } \Big) \\
\vspace{0.2cm}\displaystyle \bar D_t \rho _k+   \operatorname{div} \mathbf{j} _k  = j _k, \quad k=d,v,c,\qquad\text{and}\qquad  p(v,\eta,q _d,q_v,q _c )=p_0,
\end{array}
\right.
\end{equation} 
together with the conditions
\begin{equation}\label{thermodyn_displac} 
D_t w ^k = - \frac{\partial\mathcal{K} }{\partial \rho _k },\qquad D_t \gamma = - \frac{\partial \mathcal{K} }{\partial s}, \qquad \bar D_t\sigma = \bar D_t s+ \operatorname{div} \mathbf{j} _s . 
\end{equation} 
The first two conditions above are of fundamental importance since they indicate that the thermodynamic forces are defined from the density $\mathcal{K}$, not $\mathcal{L}$. This yields the notions of modified temperature and modified chemical potentials below.

We now specialize the system \eqref{Lagrangian_form_Eulerian_PI} to the Lagrangian density \eqref{l_moist}. Using the equalities
\[
\frac{\partial \mathcal{C} }{\partial \rho _k }= \frac{1}{ \rho }\frac{\partial p}{\partial q _k } , \quad \frac{\partial \mathcal{C} }{\partial s }= \frac{1}{ \rho }\frac{\partial p}{\partial\eta },
\]
which follows from $ \frac{\partial p}{\partial \eta } \eta + \frac{\partial p}{\partial v}v+ \sum_k \frac{\partial p}{\partial q_k}q_k=0$, using the conditions $\sum_k \mathbf{j} _k = 0$, $\sum_k j_k=0$, and the conservation of the total mass $\bar D_t \rho =0$, we finally get the system
\begin{equation}\label{PI_thermo} 
\left\{
\begin{array}{l}
\vspace{0.2cm}\displaystyle
 \rho (\partial _t \mathbf{v} + \mathbf{v} \cdot \nabla \mathbf{v} + 2 \boldsymbol{\Omega} \times \mathbf{v} )=- (\rho-\rho_0) \nabla\Phi + \operatorname{div} \boldsymbol{\sigma} ^{\rm fr}   - \nabla ( \lambda p_0 \Gamma _1 )+ \lambda \nabla p_0\\
\displaystyle\vspace{0.2cm} \Big(T+ \frac{\lambda}{\rho} \frac{\partial p}{\partial \eta }\Big)  ( \bar D_t s +\operatorname{div} \mathbf{j} _s)=  \boldsymbol{\sigma} ^{\rm fr}: \nabla \mathbf{v}  - \mathbf{j} _s \cdot\nabla  \Big(T+ \frac{\lambda}{\rho}  \frac{\partial p}{\partial\eta }\Big)\\
\vspace{0.2cm}\displaystyle\hspace{3.5cm}-\sum_k \Big[ \mathbf{j} _k\cdot \nabla    \Big( \frac{\mu _k }{m_k } - \frac{\lambda}{\rho}  \frac{\partial p}{\partial q _k } \Big)   + j_k  \Big( \frac{\mu _k }{m_k } - \frac{\lambda}{\rho}  \frac{\partial p}{\partial q _k } \Big)  \Big] \\
\vspace{0.2cm}\displaystyle \bar D_t \rho _k+   \operatorname{div} \mathbf{j} _k  = j _k, \quad k=d,v,c,\qquad\text{and}\qquad p(t, \mathbf{x} )= p_0( \mathbf{x} ),
\end{array}
\right.
\end{equation} 
where
\[
\Gamma _1 := \frac{1}{p} \Big(\sum_k \frac{\partial p}{\partial q _k }q _k + \frac{\partial p}{\partial \eta } \eta \Big)= - \frac{v }{p} \frac{\partial p}{\partial v} = \frac{ \rho c_s ^2}{p}
\]
is the first adiabatic exponent, with $ c_s ^2 = -v ^2\frac{\partial p}{\partial v}$ the squared speed of sound. This system suggests the definition of a \textit{modified temperature} and \textit{modified chemical components} as
\[
T ^\ast := T+ \frac{\lambda}{ \rho } \frac{\partial p}{\partial\eta } = T+\lambda\rho c_s^2\Gamma\quad\text{and}\quad  \frac{\mu _k ^\ast }{m_k}:= \frac{\mu _k }{m_k } - \frac{\lambda}{ \rho } \frac{\partial p}{\partial q _k }, \quad k=d,v,c.
\]
With these modified quantities, the entropy equation takes the same form as the one for the original system \eqref{system_Eulerian_moist}, namely, it reads
\begin{equation}\label{entropy_production_star} 
T^*  ( \bar D_t s +\operatorname{div} \mathbf{j} _s)=  \boldsymbol{\sigma} ^{\rm fr}: \nabla \mathbf{v}  - \mathbf{j} _s \cdot\nabla  T^*-\sum_ k \Big( \mathbf{j} _k\cdot \nabla     \frac{\mu _k ^\ast }{m_k }  + j_k  \frac{\mu _k ^\ast }{m_k }  \Big).
\end{equation}

In absence of irreversible processes, our system \eqref{PI_thermo} reduces to a multicomponent version of the pseudoincompressible system for general equations of state derived in \cite[\S4.1]{VaLeBrWoZw2013}, which itself reduces to the pseudoincompressible model of \cite{Du1989} for the ideal gas. Indeed, for the ideal gas with one component, we have $\Gamma_1=\frac{C_p}{C_v}=\gamma$ a constant and the reversible part of the right hand side of the balance of momentum in \eqref{PI_thermo} becomes
\begin{align*}
- (\rho-\rho_0)\nabla\Phi - \nabla ( \lambda p_0 \Gamma _1 )+ \lambda \nabla p_0%&=- (\rho-\rho_0)\nabla\Phi  - \rho_0\theta_0 \nabla\left(\frac{\gamma p_0 \lambda}{\rho_0\theta_0}\right)\\
&=\rho\left( \frac{\theta'}{\theta_0}\nabla\Phi-C_p\theta \nabla \pi'\right),
\end{align*}
which recovers the corresponding term in the pseudoincompressible model of \cite{Du1989}. We have denoted $\theta'=\theta-\theta_0$, resp., $\pi'=\pi-\pi_0$ the perturbations from the background potential temperature, resp., from the background Exner pressure, for an ideal gas. The relation between $\lambda$ and $\pi'$ is explicitly given by $\pi'=\frac{p_0\lambda}{C_v\rho_0\theta_0}$. The equation for the Lagrange multiplier can be obtained from the constraint and extends the Poisson equation for $\pi'$ in the pseudoincompressible model of \cite{Du1989}.

The thermodynamic consistency of system \eqref{PI_thermo} with respect to the two laws of thermodynamics will be shown below.

\paragraph{Pseudoincompressible divergence constraint.} Taking the Lagrangian time derivative $D_t$ of the constraint $p(v,\eta,q _d,q_v,q _c )=p_0$, using the entropy production equation and the continuity equations for the components $k=d,v,c$, we get
\begin{equation}\label{div_equ}
\begin{aligned} 
\nabla p_0 \cdot \mathbf{v} + p_0 \Gamma _1 \operatorname{div} \mathbf{v}  &=\frac{ \rho c_s ^2 \Gamma }{T^*} \Big( \boldsymbol{\sigma} ^{\rm fr}:\nabla \mathbf{v}-\operatorname{div}(T ^\ast \mathbf{j} _s) - \sum_k (\mathbf{j} _k \cdot \nabla \frac{\mu _k ^\ast }{m_k} + j _k\frac{\mu _k^\ast }{m_k}) \Big)\\
 & \qquad\qquad  + \sum_k \frac{1}{\rho } \frac{\partial  p }{\partial q_k}(j _k- \operatorname{div} \mathbf{j} _k ).
\end{aligned}
\end{equation}
The partial derivative in the last term is taken for the pressure expressed as a function $p=p( v, \eta ,q _k)$. Equation \eqref{div_equ} is the pseudoincompressible divergence constraint for the system \eqref{PI_thermo}. In absence of irreversible processes, and for the ideal gas with one component, we recover
\[
\nabla p_0 \cdot \mathbf{v} + p_0 \Gamma _1 \operatorname{div} \mathbf{v}=0\quad \Leftrightarrow\quad \operatorname{div}(\rho_0\theta_0\mathbf{v})=0,
\]
the pseudoincompressible divergence constraint for the ideal gas, \cite{Du1989}.

\paragraph{Energy conservation.} One of the main advantage of our variational approach is that it automatically ensures that the inclusion of the various irreversible processes respects the conservation of the total energy, i.e., the first law of thermodynamics, while this is known to be a delicate issue for pseudoincompressible approximations, see \cite{KlPa2012}. From the system \eqref{PI_thermo}, we directly compute that the total energy density $e= \rho\big( \frac{1}{2} | \mathbf{v} | ^2 + \Phi + u(1/ \rho , s/\rho ,\rho _k /m_ k)\big)$ satisfies the conservation law
\[
\bar D_t e= \operatorname{div}\Big( - (p+ \lambda p\Gamma_1) \mathbf{v} + \boldsymbol{\sigma} ^{\rm fr} \cdot \mathbf{v} -T^\ast   \mathbf{j} _s -\sum_k\frac{\mu _k ^\ast }{m_k} \mathbf{j} _k \Big),
\]
which also naturally involves the modified temperature $T^*$ and modified chemical potentials $\mu _k ^\ast$.

\paragraph{Temperature and potential temperature equations.} Thanks to the introduction of the modified temperature and modified chemical components, the temperature and potential temperature equations take the same form as in the original system, namely, we deduce from \eqref{PI_thermo} the equations
\begin{equation}\label{heat_equation_moist_PI} 
\begin{aligned}
D_t T &= - \rho c_s ^2 \Gamma  \operatorname{div} \mathbf{v}+ \frac{1}{\rho C_v ^\ast } \Big( \boldsymbol{\sigma} ^{\rm fr}:\nabla \mathbf{v}-\operatorname{div}(T ^\ast \mathbf{j} _s) - \sum_k(\mathbf{j} _k \cdot \nabla \frac{\mu _k ^\ast }{m_k} + j _k\frac{\mu _k^\ast }{m_k}) \Big) \\
&\qquad\qquad \qquad\qquad  \qquad \qquad + \sum_k \frac{1}{\rho } \frac{\partial  T }{\partial q_k}(j _k- \operatorname{div} \mathbf{j} _k )
\end{aligned}
\end{equation}
and
\[
\rho D_t \theta  = \frac{1}{ C_p ^\ast }  \frac{\partial\theta }{\partial T}\Big( \boldsymbol{\sigma} ^{\rm fr}: \nabla \mathbf{v} -\operatorname{div}(\mathbf{j} _s ^{h\ast}  ) - \sum_k (\mathbf{j} _k \cdot \nabla h _k^\ast   + j_kh _k ^\ast  )\Big) + \sum_k\frac{\partial \theta }{\partial q_k}( j _k-\operatorname{div} \mathbf{j} _k ),
\]
where $C_v^\ast = T^\ast  \frac{\partial \eta }{\partial T}(v,T,q _k)$ and $C_p^\ast = T^\ast  \frac{\partial \eta }{\partial T}(p,T,q _k)$ are the modified partial specific heat, $h _k^* =\frac{\mu _k^*}{ m_k} +T^* \eta _k$ are the modified specific enthaplies, and $ \mathbf{j} _s ^{h*} = T  ^\ast \big(\mathbf{j} _s -\sum_ k \eta _k\mathbf{j}_k \big)$ is the modified sensible heat flux, with $ \eta  _k = \frac{\partial \eta }{\partial q _k }(p,T,  q _k)$ the partial specific entropy. The partial derivative in the last term of \eqref{heat_equation_moist_PI} is taken for the temperature expressed as a function $T=T( v, \eta ,q _k)$. Note that the potential temperature in the pseudoincompressible case is defined exactly as in the original system, namely, by \eqref{def_theta_multi}, valid for any state equation.

\paragraph{Potential vorticity and circulation theorem.} From the abstract formulation \eqref{Lagrangian_form_Eulerian_PI} of our system, we can directly obtain a general form for the evolution equation for the potential vorticity $\tilde{q}$ defined by $ \rho \tilde{q} =\boldsymbol{\zeta} _a \cdot \nabla \psi  $, where $ \boldsymbol{\zeta} _a = \operatorname{curl} \left(\rho ^{-1}  \partial \mathcal{L}/\partial \mathbf{v} \right)=\operatorname{curl}(\mathbf{v}+\mathbf{R})$ and $\psi$ is a scalar field satisfying an evolution equation of the type $D_t \psi =\dot \psi$. The first equation in \eqref{Lagrangian_form_Eulerian_PI} yields 
\[
\rho D_t  \tilde{q}=\operatorname{div}\Big( \Big(\sum_k q _k \nabla  \frac{\partial \mathcal{K} }{\partial \rho_k }  + \eta  \nabla  \frac{\partial \mathcal{K} }{\partial s}  + \rho ^{-1} \operatorname{div} \boldsymbol{\sigma} ^{\rm fr}\Big) \times \nabla \psi + \boldsymbol{\zeta} _a \dot \psi \Big),
\]
which can be explicitly written by using the expression of $\mathcal{K}$ in \eqref{K} and choosing $\psi=\theta$ or $\psi=\eta$.
Similarly, Kelvin's circulation theorem for the system \eqref{Lagrangian_form_Eulerian_PI} directly follows in general form as
\[
\frac{d}{dt} \oint_{c_t} ( \mathbf{v} + \mathbf{R} ) \cdot d \mathbf{x} \oint_{c _t } \frac{1}{ \rho }\left( \sum_k\rho _k  \nabla \frac{\partial\mathcal{K} }{\partial \rho_k }+   s \nabla \frac{\partial\mathcal{K} }{\partial s} + \operatorname{div} \boldsymbol{\sigma} ^{\rm fr} \right) \cdot d \mathbf{x},
\]
where $ c _t $ is a loop advected by the wind flow.

\paragraph{Entropy production and Onsager relations.} The form \eqref{entropy_production_star} of the entropy production equation for pseudoincompressible fluids is well-adapted for an application of the Onsager relations. Namely, it suffices to consider \eqref{Onsager} with $T$ replaced by $T^*$ and $\mu_k$ replaced by $\mu_k^*$, thus giving
\begin{equation}\label{Onsager_PI}
-\begin{bmatrix}
\vspace{0.05cm}\mathbf{j} _s\\
\vspace{0.05cm}\mathbf{j} _d\\
\vspace{0.05cm}\mathbf{j} _v\\
\vspace{0.05cm}\mathbf{j} _c
\end{bmatrix}= 
\begin{bmatrix}
L _{ss} & L _{sd} & \cdots\\
L _{ds} & L _{dd} & \cdots\\
\vdots  & \vdots & \ddots
\end{bmatrix}
\begin{bmatrix}
\vspace{0.05cm}\nabla  T^*\\
\vspace{0.05cm}\nabla \frac{\mu_d^*}{m_d}\\
\vspace{0.05cm}\nabla \frac{\mu_v^*}{m_v}\\
\vspace{0.05cm}\nabla \frac{\mu_c^*}{m_c}
\end{bmatrix}, \qquad 
\begin{bmatrix}
\operatorname{Tr}\boldsymbol{\sigma} ^{\rm fr} \\
-j_d\\
-j_v\\
-j_c
\end{bmatrix}= 
\begin{bmatrix}
\mathcal{L} _{00} & \mathcal{L} _{0d}& \cdots\\
\mathcal{L} _{d0}&\mathcal{L} _{dd} &\cdots\\
\vdots & \vdots& \ddots
\end{bmatrix}
\begin{bmatrix}
\vspace{0.05cm}\frac{1}{3} \operatorname{div} \mathbf{v}  \\
\vspace{0.05cm}\frac{ \mu  _d^*}{m_d}\\
\vspace{0.05cm}\frac{ \mu  _v^*}{m_v}\\
\vspace{0.05cm}\frac{ \mu  _c^*}{m_c},
\end{bmatrix}
\end{equation}
where, according to the second law of thermodynamics, the matrices $L$ and $\mathcal{L}$ are positive. A detailed analysis of the entropy production equation for the moist air in the pseudoincompressible approximation will carried out in a future work.

\color{black}

\section{The case of the thermodynamics of the ocean}\label{3_ocean}

In this last section, we quickly indicate how to adapt the variational formulation developed above to the case of ocean thermodynamics, including the irreversible processes of viscosity, heat conduction, and salt diffusion. Sea water is a two-component system consisting of water and sea salt, with mass densities $ \rho _w$ and $ \rho _{\varsigma }$. These mass densities satisfy the continuity equations
\[
\partial _t\rho _w+ \operatorname{div}( \rho _w \mathbf{v} + \mathbf{j} _w)=0, \quad  \partial _t\rho _s+ \operatorname{div}( \rho _\varsigma\mathbf{v} + \mathbf{j} _\varsigma)=0,
\]
where the diffusion fluxes $ \mathbf{j} _w$ and $ \mathbf{j} _\varsigma $ verify $\mathbf{j} _w+\mathbf{j} _\varsigma=0$.
The equation of state of sea water is expressed in the form
\[
\rho =\rho ( \eta , T, q_\varsigma),
\]
where $ \rho = \rho _w+ \rho _\varsigma$ is the total mass density and $ q _\varsigma $ is the salinity given by $ \rho q_\varsigma=\rho _\varsigma$. Accurate approximations of the equation of state for the ocean, have been found by experiment. \textcolor{black}{The most up to date standard describing the thermodynamics of seawater can be found at \texttt{www.teos-10.org}}.

Upon using the mass densities $ \rho _w,\rho _\varsigma$ instead of $ \rho _d,\rho _v, \rho _c$, ignoring phase changes (i.e., setting $ j _k=0$), and using the expression of the internal energy $u( \eta , v, q_w, q_\varsigma)$ of sea water, our variational approach developed in Section \ref{2_moist} readily applies to ocean dynamics including the irreversible processes of viscosity, heat conduction, and salt diffusion.

In particular, with the above adaptations, the variational formulation for ocean thermodynamics is given by \eqref{VPLagr_atmosph}--\eqref{VC_atmosphere} in the Lagrangian description, and by \eqref{VP_Eulerian_atmosph}--\eqref{VC_Eulerian_atmosph} in the Eulerian description. By following the same steps as earlier, one gets, instead of equation \eqref{system_Eulerian_moist}, the system
\begin{equation}\label{system_ocean} 
\left\{
\begin{array}{l}
\vspace{0.2cm}\displaystyle
\rho ( \partial_t \mathbf{v} +\mathbf{v} \cdot \nabla \mathbf{v} +  2\boldsymbol{\Omega} \times \mathbf{v} )= - \rho \nabla \Phi- \nabla p +\operatorname{div} \boldsymbol{\sigma} ^{\rm fr}\\
\displaystyle T (\bar D_t s + \operatorname{div} \mathbf{j} _s) = \boldsymbol{\sigma} ^{\rm fr} : \nabla \mathbf{v} - \mathbf{j} _s \cdot  \nabla T - \mathbf{j} _w \cdot  \nabla \frac{\mu _w}{m_w}-\mathbf{j} _\varsigma \cdot  \nabla \frac{\mu _\varsigma}{m_\varsigma} \\
\bar D_ t\rho _w + \operatorname{div} \mathbf{j} _w=0, \quad \bar D_ t\rho _\varsigma + \operatorname{div} \mathbf{j} _\varsigma=0.
\end{array}
\right.
\end{equation}
Being derived for arbitrary state equations, our discussions of potential temperature, potential vorticity and Kelvin's circulation theorem in Section \ref{2_moist} directly apply to ocean thermodynamics.

The internal entropy production \eqref{internal_entropy_production_atmosph} takes the simpler form
\begin{equation}\label{internal_entropy_production_ocean} 
I = \frac{1}{T} \Big(\boldsymbol{\sigma} ^{\rm fr} : \nabla \mathbf{v} - \mathbf{j} _s ^h \cdot  \frac{1}{T}\nabla T -  \mathbf{j} _\varsigma \cdot  \nabla \Big( \frac{\mu _\varsigma}{m_\varsigma}- \frac{ \mu _w }{m_w} \Big)_T  \Big),
\end{equation}
with phenomenological relations
\[
\boldsymbol{\sigma} ^{\rm fr}=2 \mu  (\operatorname{Def} \mathbf{v})+ \left( \zeta - \frac{2}{3}\mu \right)(\operatorname{div} \mathbf{v} ) ,\quad -\begin{bmatrix}
\vspace{0.1cm}\mathbf{j} _s ^h \\
\mathbf{j} _\varsigma
\end{bmatrix}
=
\begin{bmatrix}
\vspace{0.1cm}A _{ss} & A _{s\varsigma}\\
A _{\varsigma s}&A _{\varsigma\varsigma}
\end{bmatrix}
\begin{bmatrix}
\vspace{0.1cm}\nabla  T /T\\
\nabla \big(\frac{\mu_\varsigma}{m_\varsigma}-\frac{\mu_w}{m_w}\big)_T
\end{bmatrix},
\]
where the coefficients verify the Onsager-Casimir relations $A_{s \varsigma}=A_{\varsigma s}$

\section{Conclusion and further directions}

In this paper, we have presented a variational derivation of the dynamics of the moist atmosphere that includes the irreversible processes of viscosity, heat conduction, diffusion, phase transition, as well as the rain process.
The variational formulation is an extension of the classical Hamilton principle for continuum mechanics, 
and is based on the introduction of new variables, called thermodynamic displacements, associated to each of the irreversible processes. We presented this principle in both the Lagrangian and Eulerian descriptions.
The impact of the irreversible \textcolor{black}{and rain} processes on the potential vorticity equation and on Kelvin's circulation theorem, was computed by staying in the general Lagrangian framework associated to our variational formulation. This provides us with a unified treatment that is potentially useful for the derivation and analysis of various approximation models for moist atmospheric thermodynamics, via the variational framework. \textcolor{black}{We illustrated this point, by deriving a pseudoincompressible model for moist atmospheric thermodynamics with general equations of state and subject to the irreversible processes of viscosity, heat conduction, diffusion, and phase transition. In particular, it showed the need to use modified thermodynamic forces, which is a priori a nontrivial step. The thermodynamical consistency of the model is automatically ensured, while this is known to be a delicate issue in general for pseudoincompressible approximations.}

\paragraph{Acknowledgments.} I am grateful to H. Yoshimura for constructive comments during the course of this work.
This work is partially supported by the ANR project GEOMFLUID, ANR-14-CE23-0002-01.

{
%\footnotesize

\bibliographystyle{new}

\end{document}